# Dr.Aid: Supporting Data-governance Rule Compliance for Decentralized Collaboration in an Automated Way

RUI ZHAO, University of Edinburgh, UK

MALCOLM ATKINSON, University of Edinburgh, UK

PETROS PAPAPANAGIOTOU, University of Edinburgh, UK

FEDERICA MAGNONI, INGV, Italy

JACQUES FLEURIOT, University of Edinburgh, UK

Collaboration across institutional boundaries is widespread and increasing today. It depends on federations sharing data that often have governance rules or external regulations restricting their use. However, the handling of data governance rules (aka. data-use policies) remains manual, time-consuming and error-prone, limiting the rate at which collaborations can form and respond to challenges and opportunities, inhibiting citizen science and reducing data providers' trust in compliance. Using an automated system to facilitate compliance handling reduces substantially the time needed for such non-mission work, thereby accelerating collaboration and improving productivity. We present a framework, Dr.Aid, that helps individuals, organisations and federations comply with data rules, using automation to track which rules are applicable as data is passed between processes and as derived data is generated. It encodes data-governance rules using a formal language and performs reasoning on multi-input-multi-output data-flow graphs in decentralised contexts. We test its power and utility by working with users performing cyclone tracking and earthquake modelling to support mitigation and emergency response. We query standard provenance traces to detach Dr.Aid from details of the tools and systems they are using, as these inevitably vary across members of a federation and through time. We evaluate the model in three aspects by encoding real-life data-use policies from diverse fields, showing its capability for real-world usage and its advantages compared with traditional frameworks. We argue that this approach will lead to more agile, more productive and more trustworthy collaborations and show that the approach can be adopted incrementally. This, in-turn, will allow more appropriate data policies to emerge opening up new forms of collaboration.

CCS Concepts: • **Security and privacy** → **Information accountability and usage control**; **Usability in security and privacy**; Privacy protections; Information flow control; *Digital rights management*; • **Information systems** → Data federation tools; • **Social and professional topics** → **Computer supported cooperative work**; *Privacy policies*; *Digital rights management*.

Authors' addresses: Rui Zhao, rui.zhao@ed.ac.uk, University of Edinburgh, Edinburgh, UK; Malcolm Atkinson, Malcolm.Atkinson@ed.ac.uk, University of Edinburgh, Edinburgh, UK; Petros Papapanagiotou, pe.p@ed.ac.uk, University of Edinburgh, Edinburgh, UK; Federica Magnoni, federica.magnoni@ingv.it, INGV, Rome, Italy; Jacques Fleuriot, jdf@inf.ed.ac.uk, University of Edinburgh, Edinburgh, UK.

# 1 INTRODUCTION

Collaboration across institutional and discipline boundaries is an increasing practice in research today, whether through tight alliances or loosely coupled federations. In these collaborations, data sharing is a core activity, combined with analysis and modelling computations. There are initiatives such as (linked) open data [71], Research Objects [27] or FAIR [75] to provoke data sharing to wider audiences, to improve reproducibility, to broaden impact, etc. However, in many cases, data providers or governors need to establish and extend data governance rules, due to governmental policies or properties of the data (e.g. containing sensitive information) [52]. In such circumstances, it is impermissible to simply make it "open data". Current practice for such situations often requires data users to submit applications and undergo training on security, privacy, sensitivity[1] and ethical data management before gaining access to the data, and their results may also require to be screened before they are allowed to disclose them to a wider audience. Policing such systems is onerous for data providers and compliance is tedious and time consuming for researchers. It may inhibit research even when the restrictions only pertain to a small portion of the data. One can easily observe the widespread polarization of data-governance practice, which we believe is due to lack of automated help for sophisticated data-use rule specification and compliance.

This socio-technical problem of data sharing, reusing and research reproducibility has been recognized by previous work in Computer-Supported Cooperative Work (CSCW) and related fields. They discussed the necessity and benefits of data and software sharing [32, 39, 77], the keypoints and burdens for implementing that [33, 58, 76], and how that is perceived and expected by the end users [20]. They provide useful insights and discoveries, such as the necessity of extra data-use policies during data sharing (e.g. prevention of scooping [58]). But the lack of sufficient methods, especially systems, for dealing with data-use policies, remains a critical unsolved issue.

Taking a broader view, this issue applies beyond traditional research data. It covers the technologies and methods to protect privacy and promote reproducibility in non-traditional data, e.g. social media [40], the discussion about the issues in traditional consent-based user agreement [47, 59], and emerging issues for IoT (Internet of Things) or smart devices [20, 68, 79]. They all pose data-use rule specification and compliance challenges with non-centralized data processing. Therefore, overcoming this problem requires improving the technology we have, to enable systematic monitoring and enforcement of data use policies, while gradually shifting the social practice. In the end, a new paradigm of computer-supported rule formulation and compliance will emerge to facilitate cooperative and collaborative work.

Facing the necessity of better methods to deal with data-use policies, different approaches try to tackle this issue with different viewpoints and goals (Section 2), including jurisdictional constraints and automated frameworks with formal models constraining data-use. Although different perspectives have different features and focuses, there are two major reasons for using a formal model: (1) to avoid the ambiguity in natural languages; (2) to expose/extract the similarity in data-governance rules, despite their representational heterogeneity. Figure 1 presents rules on data-governance selected from public online sources (usually named "Terms of Use"). It highlights the most informative parts. It can be noticed that most parts are less informative or even unimportant to the policies themselves. Using a formal model can reduce such issues, and make the rules concise and accurate.

In our research, we take account of this data-governance context: data are from different sources and are processed by different bodies; the data processors are in different institutions who may not have tight collaboration agreements. Output data can be taken as input for other work, immediately as part of a current campaign, or in a currently unplanned future campaign involving different

---

[1]Sensitive encompasses personal data, commercial-in-confidence and content such as emergency-response locations to avoid panic and media.

I agree to restrict my use of CORDEX model output for non-commercial research and educational purposes only. [1]

In publications that rely on the CORDEX model output, I will appropriately credit the data providers by an acknowledgement similar to the following: "We acknowledge..." [1]

You may extract, download, and make copies of the data contained in the Datasets, and you may share that data with third parties according to these terms of use. [2]

When sharing or facilitating access to the Datasets, you agree to include the same acknowledgment requirement in any sub-licenses of the data that you grant, and a requirement that any sub-licensees do the same. [2]

Data is non-transferrable (other than as permitted in the licence) and confidential in nature. [3]

Data is not to be used to identify, contact or target patients or general medical practitioners. [3]

[1] CORDEX terms of use: https://www.hereon.de/imperia/md/assets/clm/cordex_terms_of_use.pdf
[2] World Bank Terms of Use for Datasets: https://www.worldbank.org/en/about/legal/terms-of-use-for-datasets
[3] CPRD client application form: https://www.cprd.com/Data-access

Fig. 1. Highlighting important terms to be encoded in a sample of data-governance rules from three sources

partners. We call this a *federated data-processing context*. Such a context is aligned with data-intensive research [38], where data have a wide-range of different governance requirements (policies). Collaboration across institutional boundaries is common practice for such a context. It is essential to properly comply with the data-governance rules, otherwise a collaboration may collapse and future collaborations with the same partners may become unachievable. Here we present an example scenario extracted from research practice (also in Figure 2):

> *Dataset (D) comes from a data provider (DP). It contains some sensitive information in its column "*DoB*" (Date of Birth). The rest of the data is not sensitive. DP wants to be informed of all uses of the* DoB *column, to prevent harmful disclosure. Apart from that proviso, DP permits the data to be shared with the public, and allows anyone to produce derived work. DP also wants to be credited for producing this dataset by being cited in publications produced by users. Therefore, they state these two requirements in their data-use policy, and have set up a use-reporting mechanism (*report.example.ac*). A data user UA processes the data and produces two output datasets: DB and DC. DB does not contain* DoB*. DC contains only the* YroB *(Year of Birth) derived from* DoB*. UA wishes to share these two datasets with other researchers.*

Naturally, a few consequences emerge from this scenario. After obtaining the data D, the data user UA performs data processing independently from DP. As specified in the example, DB does not contain the sensitive information DoB, so it would not be bound to the obligation of reporting uses; DC contains derived information YroB so it *can* be considered as still bound to the reporting obligation. Therefore, when UA shares DB and DC to other, appropriate policies for each of them (derived from the original policy) should be attached as well. Similarly, any future users (e.g. UX) using DB is not bound to the reporting obligation too, while users of DC are. If a user UY uses both DB and DC, he/she is still bound to all the original policies (union of the policies of DB and DC), even though UY obtains data from UA instead of DP and might not be aware of the existence of DP. Similar to UA, UX and UY can perform arbitrary processing on the data, creating different consequences for the policies.

From that, we can identify 5 major properties, which are also issues to solve in such contexts, below:

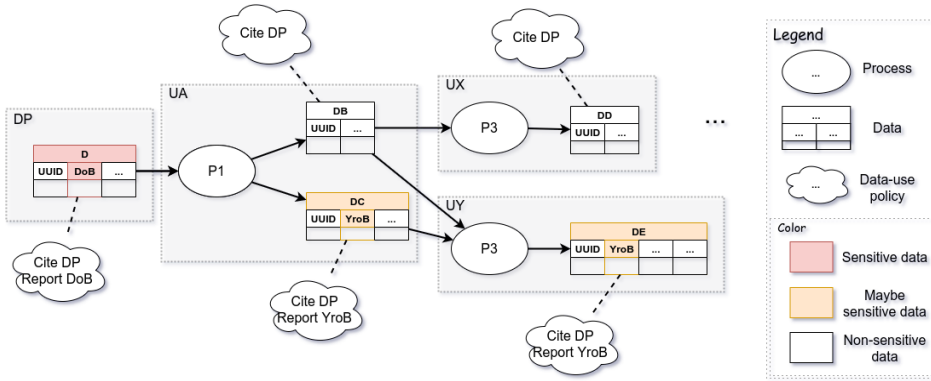

Fig. 2. Illustration of the example use case. Each user has their governance boundary, and are subject to the data-use policy from the data they use.

†1 (Personnel) **Scattering**: data processing is multi-institutional so that data providers and data processors are rarely in the same institutional framework.

†2 (Rule) **Propagation**: derived data (output data) can be used as input data further, by the same or different people in the current activity or some future activity.

†3 (Rule) **Diversity**: policies not only impose access control, but also contain general *obligations* that current and future users should fulfil.

†4 **Dynamic** (rule) **application**: processes change data and therefore can revise / change the policies applied to data, in particular lowering the policy restrictions.

†5 (Rule) **Combination and separation**: processes can be multi-input-multi-output (MIMO). This may also be checked in two halves:

  †5.1 (Rule) **Combination**: processes may take multiple inputs with different policies.

  †5.2 (Rule) **Separation**: processes may produce multiple outputs with different policies.

These identified issues demonstrate the necessity of having automated frameworks to support both the data providers and the data users to deal with rules. Section 2.2 summarizes the different features and focuses on related research taking a similar direction, and concludes that there is a lack of frameworks to solve all identified issues in federated contexts.

Therefore, in this paper, we present an intelligent framework called Dr.Aid (Data Rule Aid), supporting reasoning about derivation of data-use policies and checking compliance status, which addresses all these aspects and therefore supports data-use rule compliance in a broad range of federated contexts. In particular, this framework handles data-use policies in MIMO data-flow graphs, which we have not found elsewhere. It also provides an extensible language including obligations, which are also not well-supported in existing frameworks. We envision this framework providing a foundation for a future with automation supporting data-use policies, initiating more productive and sustainable data-intensive research collaborations.

A broader background and related work are presented in Section 2. The introduction to the framework is in Section 3. We present the evaluation in Section 4. In Section 5, we discuss the current limitations and future work; finally, in Section 6, the conclusions are drawn. The appendices contain additional details referenced in the main text (e.g. encodings of the data-use policies). Complex figures and longer listings are also in the appendices.

## 2 BACKGROUND AND RELATED RESEARCH

This section discusses the background that shapes our research goals and reviews related work with similar goals.

### 2.1 Background

Processing data with the support of computer systems is one of the most common collaboration practices today, particularly for research. This is often denoted as data-intensive research [38], where the role of data sharing is dominant.

The importance of data governance, data ethics and privacy has risen in recent years driven by the widespread application of machine learning [50] and the Internet of Things (IoT) [51, 79], which generate and use massive amounts of data on a daily basis. This concurs with the so-called "biggest lie on the Internet" [59] i.e. the fact that most people explicitly accept website Terms of Service and Privacy Policies without reading or understanding them. This raises the same issues and modes of failure; whenever people try to enhance their control over data usage, problems arise due information overload. Legislative approaches, such as the European General Data Protection Regulations (GDPR), bring some consistency and return control back to the data subject (normally the user) [1], but they do not eliminate the complexity for people, leaving them facing the issues of finding, understanding and complying with data rules unaided. Therefore, appropriate methods and practical frameworks are needed to facilitate every stakeholders' role relating to data ethics and governance.

Efforts have been made to address challenges around privacy by algorithmically eliminating the necessity for and the use of original sensitive data, namely differential privacy [19] (where sensitive-data details are obscured in synthesised derivatives) and federated learning [50] (where sensitive data are restricted to local processing). They provide useful methods for protecting privacy while also keeping high accuracy and personalization. However, issues remain because privacy is not the only element for data governance and ethics. Besides, in many cases, sharing sensitive data is necessary and desired [46], so that decentralized and fine-grained governance is explicitly required.

Some research points out the diversity of people's preferences, and provides automated agents to negotiate with the data accessing body on behalf of the user [25, 42]. This directly addresses the governance and ethics challenges with reduced human effort, particularly in the context of IoT and smart devices with unpredictably many negotiation/authorization requirements. However, they follow a traditional view of data processing where the data is used in one processing step (directly by the organisation to which consent has been granted) or in a limited step by an authorised third-party. Data processing has no context and one consent governs data usage forever; derived data products are beyond the scope of control of the consent. In typical general contexts, data processing can be multi-staged and/or conducted by multiple bodies, thereby exposing the limitations of these solutions.

As we describe below (Section 2.2), other research focuses on distinct policy requirements and the use of automated frameworks to check and/or ensure compliance. This can reduce effort (for data consumers), facilitate the authoring and maintenance of data governance rules (for data providers), and maintain compliance for not only the initial data but also its derivatives. We view this as a necessary direction, such that it may be combined with the automated negotiation agents described above to enable full-fledged practical frameworks maximizing social benefits while also respecting individuals' rights and preferences.

Table 1. Summary of framework features regarding our identified issues for realistic contexts where multiple distributed participants progressively import, combine and process data.
✓ means supports; ✗ means does not support; ✓ means partially (often very limitedly) supports; ? means unknown.

| Framework | Scattering | Propagation | Diversity | Dynamic application | Combination | Separation |
|---|---|---|---|---|---|---|
| E-P3P[43] | ✗ | ✗ | ✓ | ✗ | ✗ | ✗ |
| Thoth[29] | ✗ | ✓ | ✗ | ✓[2] | ?[3] | ✗ |
| DAPRECO[26, 66] | ? | ✗ | ✓ | ✗ | ✗ | ✗ |
| Smart object[67] | ✓ | ✓ | ✓ | ✗ | ✓ | ✗ |
| CamFlow[62] | ✓ | ✓ | ✓[4] | ✓ | ✗ | ✗ |
| Meta-code[41] | ✓[5] | ✓ | ✓[6] | ✓ | ✗ | ✗ |
| Dr.Aid [our work] | ✓ | ✓ | ✓ | ✓ | ✓ | ✓ |

## 2.2 Related research

In this part, we discuss the related research that also investigates automated systems to ensure compliance with data governance rules. We discuss them below, and summarize their achievements relative to our five identified issues in Table 1.

One direction of research focuses on ensuring the compliance in a known closed context (e.g. within an institutional boundary). For instance, E-P3P [43] provides a formal model to check compliance before granting access to data. It also introduced the concept of *sticky policy* [56] (see below). Thoth [29] uses a more flexible logic-based formalisation to encode access control rules as well as automatic declassification conditions, but can not describe *obligations* (required actions as a consequence of using the data) as [43] does. DAPRECO [26, 66] is a legal-modelling approach taking a similar view, converting legal documents (e.g. EU GDPR, General Data Protection Regulation) to logical expressions and checking compliance of some processing. These approaches have different strengths and flexibility, but they hold a narrow view of data processing: "that data processing seldom affects the applicability of policies". As a result the data-use policy for the input data invariably pertains to all derivatives until the result meets the *declassification* requirement specified by the original policy maker / data governor; the declassification makes the result no longer bound to the original policies, nor to any policies. Sticky policy [56, 63] raised the policy enforcement issue for decentralized contexts, and provided a conceptual framework for maintaining policy compliance in such contexts. [67] (denoted as *smart objects*) provides a model to encode not only the direct data-use policy, but also the mechanism to derive the policies for derived data. Such frameworks are aware of the decentralized context and provide rich controlling power to the data provider. But they require a close collaboration between the data providers and the data users to allow data providers to foresee the processes that the data may go through and encode that in the policies. As a summary, these research constitute useful approaches when the data providers and the data users collaborate closely or are within the same institutional framework. But for loosely coupled contexts (such as the federated context identified above), it is almost impossible to predetermine the processes the data will go through as methods evolve during the collaboration, and therefore such frameworks could not provide the required support.

---

[2] It has the *declassification* rules, but it cannot model processes.
[3] A figure in the paper seems to imply this, but it did not discuss this.
[4] Only the *integrity* label.
[5] Although meta-code allows processes to change data policies, the approach is highly centralized by using role labels.
[6] Through meta-code, custom arbitrary program code.

Aside from frameworks, there are dedicated policy languages, such as the Open Digital Rights Language (ODRL) [18] and the eXtensible Access Control Markup Language (XACML) [17]. XACML is an XML-based standard used to describe access control; ODRL is a W3C standard based on semantic technologies to describe various aspects of data's terms of use. Regardless of their differences, the primary purpose of these languages is to formally represent data-use policies and check whether a *single* use conforms to them. They do not address rule propagation, data derivation, merging or separation. Thus they face the same issues as the frameworks discussed above.

A few other researchers address the propagation and dynamic application issues, explicitly focusing on allowing processes to change the policies associated with derived data. Meta-code [41] and CamFlow [62] utilize concepts from (decentralized) Information Flow Control (IFC) [57] as the foundation to specify the policies and change of policies, and make different extensions. The basic concept is to assign tags to data and to specify additional constraints as tags to processes/programs: processes have specified input-tag compatibility, so only compatible data can be permitted as input; processes will produce output, so the tags for output data are specified along with the processes. Meta-code [41] introduced the *meta-code* concept to model the policies that can not be captured by role tags, using custom program code; CamFlow uses the model of decentralized IFC with two labels (each contains a set of tags), secrecy and integrity, to represent different policy semantics; output policy is specified by manipulating input labels for each process, called *label change*. As a result, Meta-code supports richer types of policies but lacks formalization, making static analysis difficult. CamFlow has semantics limited to the two labels, within access control. Both approaches and research building on them that we have found do not support MIMO processes.

Our framework, Dr.Aid, provides a solution for all five issues, in particular the rule diversity with support for MIMO processes (see Section 4.3 for an in-depth comparison). Its language model is derived from our previous work [78], which consists of the *data rule* part to model the data-use policies and the *flow rule* part to model the change of data rules on a process. It supports obligations (actions to be performed after using the data), which makes it different from other traditional solutions that focus on access controls. We consider it closely related to but different from the decentralized IFC model used in CamFlow (denoted as DIFC below): On the one hand, they both separate the definition of data-use policies and the process policies which manipulates the data-use policies; on the other hand, DIFC binds semantic meanings to the tags and manipulates tags, while our model separates the manipulated element (the *attributes*) and the semantic element (the *obligations*). This results in a more flexible and semantically extensible language. The original model in [78] is limited in several aspects and lacks a formal underpinning. Therefore, we provide a revised version of it currently operational in Dr.Aid, which we introduce in the following section.

## 3 THE DR.AID FRAMEWORK

With the brief description of the general language model used in Dr.Aid (Data Rule Aid) above, the general outline of the Dr.Aid framework can be perceived – coupling data flow with rule flow, addressing the MIMO requirements and supporting dynamic rule application, as illustrated in Figure 3. This section presents the more complete design of the Dr.Aid framework, including the representation of rules and the architecture that tracks their applicability.

As mentioned above, the language model used in Dr.Aid derives from our previous work [78]. The key difference is to extend the generality of the model. We constructed an implementation using the revised language. The significant features are:

- We developed a formal description of the model.

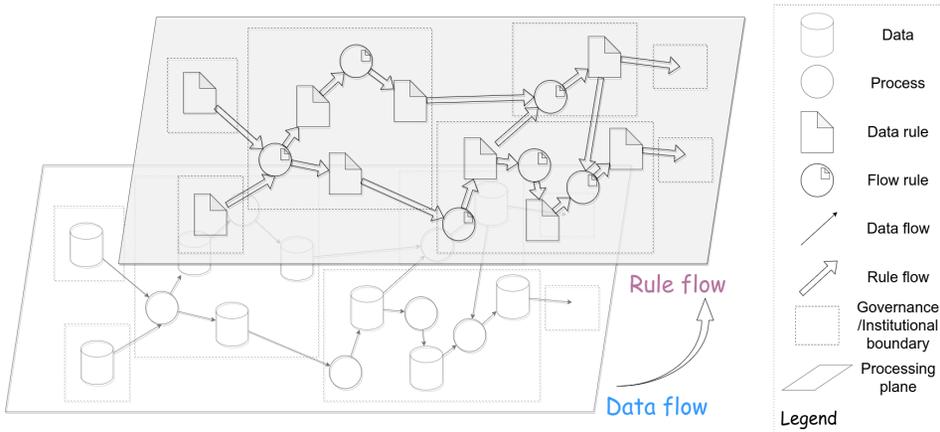

Fig. 3. Conceptual design of the Dr.Aid framework, shifting from the *data flow* at the lower level to the *rule flow* at the upper level

- We created a logical interpretation of the model using a well-studied logic system, namely situation calculus [54, 65]. This supports whole-graph reasoning (as opposed to process-by-process reasoning).
- We integrated our implementation with a well-known dedicated situation calculus reasoner, Golog [49].
- We provided a more flexible model with slots for the activation conditions;
- We provided an abstract intermediate graph model to support compliance checking from both data-streaming (S-Prov[7] for dispel4py [34]) and file-oriented (CWLProv[8] for CWL[21]) workflow systems.
- We co-developed and evaluated the model and framework with real-life use cases (in Section 4).

## 3.1 Context and assumption

In the data-intensive research context, researchers usually do not generate initial input data themselves. Instead, they use data from different upstream data sources, which can be automated systems (e.g. sensors) or data generated by other researchers. When they do generate input data, there are usually established widely-acknowledged procedures stipulating how it is authorised for further use specifying any restrictions on that use. Therefore, we can assume a standard protocol which associates data-use policies with the data. In most cases, researchers do not need to *provide* the policies, they or the process selects and paramaterizes an existing one.

When designing Dr.Aid, we assume that a suitably populated repository or catalogue of data and workflow processes (not necessarily workflows) will eventually exist (see Section 6). It will provide computer-interpretable information associating formal rules with data, software and resources, and propagation rules with processes. We observe this often unstated assumption in contemporary research. We consider this a fair assumption for our scenario, not only because related research assumes this, but also because there are dedicated work on cataloguing workflows [36, 53], data [72], and surrounding information [44]. Manual exploration and reusing may be tedious and overwhelming, but it can be more feasible with the help of automatic mining, matching and composition [35, 37]. As we will show later in this section, we identify processes from the provenance, which can be traced back to their original definition in the workflow. A similar procedure examining the import parameters and direct input information defining data sources does the same for data. With such a repository, one can automatically obtain the associated data rules and flow rules for the data and processes. In a deployed system this will depend on a standard protocol to establish and retrieve this information.

The example described in Section 1 will be used. More specifically, we assume a user UA uses a process which produces dataset DB from output port[9] output1 and dataset DC from output port output2.

## 3.2   Design and language

This part introduces the language model: the *data rule*, the *flow rule*, and how they interoperate.

### 3.2.1   Data rule.
The *data rules*, as a means to model data-governance rules, are associated with data – each data object is associated with a data rule set. They contain two main building blocks: *attributes* and *obligations*.

*Attribute.* An *attribute* describes properties of the data and is represented as a triple (N, T, V) of a name N, a type T and a value V. It is the main building block in the data rule, which is used by obligations, and receives special attention in *flow rules*: as will be described below, the refinements in flow rules manipulate attributes, similar to the label change in DIFC; when an attribute gets removed, all obligations bound to it are removed too. This presents a mechanism allowing processes to change the data rules associated with the outputs, leading to a decentralized manipulation mechanism of data-use policies.

*Obligation.* An *obligation* specifies an action (to be performed by the user) that is triggered under specific conditions, as well as its "dependency" attributes. Formally, an *obligation* is a triple (OD, VB, AC) consisting of an obligation definition OD (the action to perform upon activation), a validity binding set VB (additional applicability constraints), and an activation condition AC (the triggering condition). The OD is another tuple (OA, AR) where OA is the obligated action class and AR is a list representing action parameters. In particular, each element of AR and VB refers to an attribute in the data rule, which forms a binding of the attribute (see *flow rules* below for details). The activation condition AC is a boolean expression, which will be evaluated into true or false with runtime information when checking the activation of obligations (Section 3.3.1). Appendix A summarizes the available *slots* which are the aspects that can be checked (e.g. process type, time of execution, etc).

*Example encoding.* For instance, the example rule above regarding the reporting of any use of the sensitive "DoB" field to an example URL `report.example.ac` can be modelled as follows:

$$attribute(pf, column \text{ "DoB"})$$
$$attribute(ru, url \text{ "report.example.ac"})$$
$$obligation(report\ ru, [pf], action = *)$$

---

[9]Processes, the main building blocks of scientific workflows, can take multiple inputs and multiple outputs, each through one of its input ports and output ports.

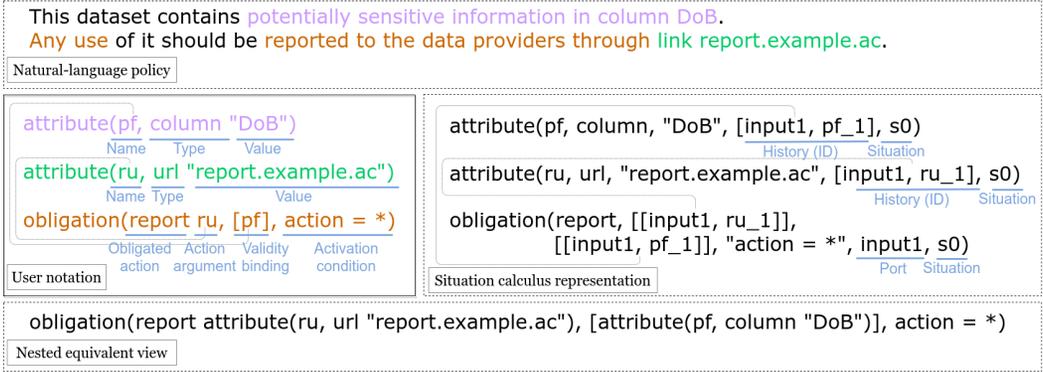

Fig. 4. Encoding (and equivalences) of the example data-governance rule (associated with input1)

Most elements in this formal notation, which we call the "*user notation*" can be directly mapped from the original natural language rules. This is further converted to the "*situation calculus representation*" automatically to include additional information used by the situation calculus reasoner during inference (see Section 3.3.3). Figure 4 shows a comparison between different representations of this rule. The modelling shall be explained as: the rule segment column"DoB" is modelled as an attribute whose type is column, value is DoB, and name is pf (*private field*); the rule segment url report.example.ac is modelled as another attribute whose type is url, value is report.example.ac, and name is ru (*report url*); the main content is an obligation declaration with reference to these two attributes, whose obligated action is report, action argument is ru (referencing the ru attribute), validity binding is a list with one element [pf] (referencing the pf attribute), and activation condition is action = * meaning it would activate when the data goes through a process with *any* action type. We can see that the necessary information in the natural-language policy has been encoded in the formal notation.

*3.2.2 Flow rule.* The *flow rules*, on the other hand, describe how the data rules would flow through a process, reflecting the underlying data propagation and processing. They involve three types of actions: *propagate*, *edit*, and *delete*.

*Propagate.* *Propagate* specifies the general flow of data rules from input ports to output ports, when no edit or delete is applied. It is a tuple $pr(P_{in}, P_{out})$ where $P_{in}$ is the input port to propagate data rules from and $P_{out}$ is an output port to propagate data rules to. A shorthand $pr(P_{in}, Ps_{out})$ is used to specify multiple output ports (a list of output ports $Ps_{out}$) for the same input port $P_{in}$.

*Refinements − edit & delete.* After specifying propagation, further refinements can be done to the data rules, to reflect the processing and modification of underlying data, specifically the **edit** action and the **delete** action. *Delete* is specified as $delete(P_{in}, P_{out}, N, T, V)$ where $P_{in}$ is an input port, $P_{out}$ is an output port, N is the name of a attribute, T is the type of an attribute and V is the value of an attribute. It acts as a *filter* to match all the data rules (of the process), and remove every matched *attribute*. As a consequence, every *obligation* which refers to these *attributes* (in their action parameters or validity bindings) is removed as well. Similar to delete, *edit* is specified as $edit(P_{in}, P_{out}, N, T, V, T_{new}, V_{new})$ where $P_{in}, P_{out}, N, T$ and V are the same as those in *delete*, $T_{new}$ is the new type of the attribute and $V_{new}$ is the new value of the attribute. The filter is similar to delete, but the matching attributes will have their type and value updated to the specified new type $T_{new}$ and new value $V_{new}$. In addition, each field of the filter (excluding new values) can be specified as a special value ∗, which corresponds to *any possible* value.

*Example encoding.* For instance, the flow rule for the example process can be specified as (in the user notation):

$$pr(input1, [output1, output2])$$
$$delete(input1, output1, ∗, column, "DoB")$$
$$edit(input1, output2, ∗, column, "DoB", column, "YroB")$$

This says the data rules will be propagated from input1 to both output1 and output2, under revision a) to delete attributes from port input1 to port output1 with *any* (∗) name, type column and value "DoB", b) to change attributes from port input1 to port output2 with *any* (∗) name, type column and value "DoB" to type column and value "YroB". By definition of the semantics, the revision a) also deletes any obligations bound to the deleted attributes from output1, i.e. the reporting obligation, but it won't affect output2.

## 3.3 Reasoning mechanism

Reasoning is performed by taking the data rules for each input port, executing flow rules, and obtaining the data rules for each output port.

Using the example with the encoding above, the outputs can be automatically calculated to have the following data rules:

*Data rules of* output1 *(i.e. of DB).*

$$attribute(ru, url "report.example.ac")$$

*Data rules of* output2 *(i.e. of DC).*

$$attribute(pf, column "YroB")$$
$$attribute(ru, url "report.example.ac")$$
$$obligation(report ru, [pf], action = ∗)$$

Note the dangling attribute ru from output1 is deliberately kept by the semantics. This design considers the accreditation needs of data providers to leave information in the data rules, and also keeps the language specification simple. While other researchers may prefer to prune the dangling attributes for the sake of simplicity in the data rules, we argue that this is not critical and is merely a design choice.

The reasoning process is intuitive. As demonstrated above, the data rules come in from some input port, which is attached to them during reasoning as necessary information for flow rules; when there are *propagate* rules, the corresponding output ports are associated too, so the *edit* and *delete* can be carried out; after the flow rule processing, the resulting data rules are sent out through the corresponding output ports.

*3.3.1 Obligation activation.* The procedure above allows us to derive successor data rules. Further reasoning allows checking the activation of obligations. This is done by checking the activation condition of the corresponding data rules at the beginning of each process using contextual information. For the obligations whose activation condition is evaluated to true, their obligation declarations OD (including the referenced attributes) will be extracted, and will be put into a separate storage in our implementation. The applied contextual information contains the process information (e.g. process type), the execution information (e.g. the stage during execution) and the provenance information (e.g. the user), as summarized in Appendix A.

*3.3.2  Merging and deduplication.* Through the flow rules, the rule merging and separation issues is mostly solved – the user is able to explicitly specify how the rules would flow. However, there is still an undiscussed case when different incoming data rules have duplicated entries. Consequently, the output data rules may have duplicated entries propagate (as-is or as the result of editing) if handled naively. Logically, the data rules coming from and going to a port form a set. Therefore, when merging happens, the framework also removes duplicated entries.

*3.3.3  Situation calculus formalization.* In our work, the language and the reasoning mechanism is provided with a logical background using situation calculus [55], a well-studied logical formalism to characterize dynamic domains, consisting of an extension to first-order logic. Based on these facts, situation calculus is both simple and a good fit for our requirements.

Our method is to align the model components and reasoning with the constructs in situation calculus, which is to model the data rules (plus the associated ports) as *fluents*, the flow rules as *actions*, the different steps of flow rule execution as *situations*, and the reasoning as the *projection task*, i.e. given a target situation (state) $S_f$, query the fluents that hold in $S_f$.

The fluent-based situation calculus representation, as shown in Figure 4, contains information about the *history*, i.e. the ports that the information has gone through in each stage, and the current *situation*, i.e. the current state in addition to the parameters of the formal specification discussed earlier.

Due to the particular focus and length consideration of this paper, we do not present the full explanation of this formalization. See Appendix B for the list of relevant axioms (precondition axioms and successor-state axioms).

## 3.4  System implementation

We built a system implementing the reasoning mechanism above, as well as reading and handling other relevant information[10]. The system is mainly implemented in Python and uses Golog [49] (on SWI-Prolog)[11] as the situation calculus reasoner; it uses Owlready2 [48] for ontology-related operations. Figure 5 gives a high-level view to the architecture of our implementation. The main goal of the system is to take a data-flow graph whose input data and processes have rules (data rules and flow rules) associated, and to perform reasoning over the data-flow graph to obtain: (1) any activated obligations; (2) data rules associated with output data after the processing. Therefore, in turn, the obtained derived data rules can be used as input data rules for further reasoning.

*3.4.1  Input source.* The implementation performs retrospective analysis by taking provenance traces as the source of data-flow graphs. The main benefit of provenance is that it allows us to abstract from the implementation details of various (workflow) execution systems, thanks to the standard core ontology, W3C PROV-O [73], the interoperability provided by the semantic technology, and the standard query language SPARQL [74].

Our system uses provenance traces produced by scientific workflows [22], which have two major types, file-oriented and data-streaming. For file-oriented workflow systems, each process takes inputs from data files, and produces outputs to data files. The files are either hard-coded in the source code, or passed in as parameters to the processes. On the other hand, processes in data-streaming systems read inputs directly from the outputs of its predecessor processes, without storing to files. The outputs are usually small data units, each representing a meaningful segment of the full output (e.g. a line in a table, a number in a sequence, etc). Such differences give them different capabilities, and also imposes different requirements for the provenance scheme. Because PROV-O is a low-level model, extensions are developed to provide higher-level descriptions for specific needs. In our implementation, we support two provenance schemes for each one of them, namely CWLProv and S-Prov, as illustrated above.

---

[10]Source code can be obtained from https://github.com/renyuneyun/dr.aid/tree/cscw2021
[11]The Golog implementation is obtained from http://www.cs.toronto.edu/cogrobo/main/systems/index.html.

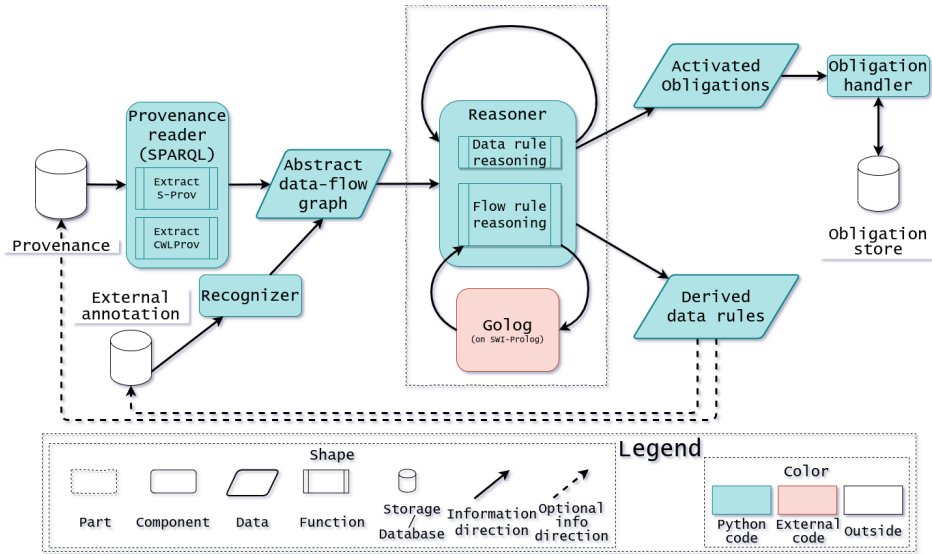

Fig. 5. High-level view of the system architecture

In order to support the distinct properties of different schemes, Dr.Aid uses an abstract intermediate representation for the data-flow graph (a visualization example can be found in Figure 6), through SPARQL queries. The main reason we don't use PROV-O directly is because PROV-O is too low-level and causes redundancy in the data production and consumption for data-streaming workflows (S-Prov in our example). In addition, PROV-O is retrospective while our model is not; PROV-O implies the strict existence of intermediate *entity* (e.g. data) between two *activities* (e.g. processes), which can become a limitation in the future to expand the use cases to process graphs without explicit data, e.g. BPMN [60].

*3.4.2 Recognizer module.* In order to associate rules with the data-flow graphs to cope with the fact that not all data and processes have rules associated with them already, we use the *recognizer* module. Before reasoning, the recognizer checks the data-flow graph, finds matching rules from its database, and injects these extra rules to the data-flow graph. The recognizer also supports identifying processes that need to add additional rules apart from its inputs (e.g. those downloads data internally with no input ports), and inject data rules to such processes. In our implementation, the database is stored as a JSON file.

The database used by the recognizer can also be used to store the reasoning results, i.e. data rules associated with the output data. This is useful for doing experiments, and also useful when the provenance store does not allow to write back (e.g. due to permission issues).

*3.4.3 User actions as virtual processes.* Inspired by PROV-O, Dr.Aid uniformly treats user actions and computational processes. Therefore, user actions can be injected as *virtual processes*, and the reasoning will go through the same procedure to check activation and/or propagate data rules. In our implementation, this is done by adding extra annotations to the abstract intermediate graph representation to include virtual processes when instructed.

*3.4.4  User interaction.* The implementation has two major user interaction points: (1) Setting the data (provenance and rules) source and execute the reasoning; (2) Checking the activated obligations. Both points are explained above, while the 2nd point is only briefly explained when introducing the activation of obligations. The users are expected to check the activated obligations after the reasoning, and perform actions accordingly. This is enough for experimental purposes as proof-of-concept. In an ideal situation, the 1st point can be automatically completed, with the help of workflow/data repositories/catalogues, and the users are expected to check only the 2nd point, through a proper notification mechanism.

## 4  EVALUATION

In this section, we present the evaluation we performed for Dr.Aid. The evaluation covers:

(1) the ability of our implementation to handle real-world data-flow graphs in collaboration contexts;
(2) its advantage against other frameworks;
(3) the capability of the language for expressing real-world data-governance rules.

Our first evaluation is based on the use of Dr.Aid in two real-life scientific workflows: cyclone tracking for global-warming impact modelling and Moment Tensor in 3D (MT3D) computing the expected impact of an earthquake. The second evaluation is based on the scenario extracted from real-world research practice, described previously in Sections 1 and 3. Then we evaluate the capability of the language to specify a selection of diverse real-world published data-governance rules.

### 4.1  Experimental consistency

Each evaluation has specific properties, but there are commonalities shared between them (particularly the 1st and the 3rd). The most important one is the procedure to convert from natural-language policies to the formal representation. We have standardised the procedure for this:

(1) Identify and obtain nested rules if any;
(2) Remove unnecessary information from the rules;
(3) Identify *actioning* rules, in particular obligations;
(4) Find the terms in the rules that identify the data or critical properties of data that need to be carried with data, as attributes;
(5) Identify *implied* rules;
(6) Write in the user notation where possible;

The *actioning* rules are the rules that describe an action, which can be an action/behaviour to be complied with when using the data, an action to be performed after using the data, or an action imposed by someone else (usually the data provider) on the user. They are the major contents of rules to be encoded in our model. The *implied* rules are implicit in our model and need not be encoded. An example is *"the user is allowed to redistribute the derived data"*. Implicit behaviours can be explicitly overridden when necessary.

It is worth noting that not every sentence in the natural-language policies can be modelled using our formal language, because those sentences describe contextual information, or because they are beyond the capability of our current model. We discuss such cases as they arise.

Therefore, following this standardized encoding procedure, we measure its effect using the following information:

(1) The total number of sentences in the original natural-language (English) policy;
(2) The total number of rules in the original policy;
(3) The total number of actioning rules;

(4) The total number of implicit rules;

(5) The total number of encoded rules.

## 4.2 Framework evaluation

The framework evaluation tests the capability of the whole framework with use cases that involve typical collaborative use of data and computational methods for global research addressing environmental hazards [24, 45]. It considers the language encoding, the system implementation, the extracted information, the reasoning result, etc.

As mentioned previously, the selected instances of collaborative behaviour are climate-scientists setting up and running *cyclone tracking* workflows and seismologists setting up and steering workflows to estimate an earthquake's impact in an area they select, either to advise emergency response or to improve regional models for future use (*MT3D*). We use the provenance traces generated by the executions of these workflows, and encode the data-use policies of the data selected by users and imported from an open-ended set of providers during these executions. The information originates from the scientific researchers who authored and executed these workflows; we collected, transformed and analyzed them (e.g. traced the data-use policies and encoded them); finally we consulted these researchers to validate our results as expert opinions, which includes whether the collected policies are complete, whether the encodings reflect the expected meanings of the original policies, whether the extracted data-flow graph matches the workflow specification, whether the derived rules match their expectation, etc.

As well as being typical of the collaborative use of data in multi-disciplinary, multi-site loosely coupled federations, we choose these two examples because they contain complex data use patterns, involving multiple processing stages, data separation and data merging. They also illustrate Dr.Aid's applicability for different types of workflow systems, data-streaming with `dispel4py` and task-oriented with CWL. The MT3D workflow consists of multiple sub-workflows set up and individually steered by the seismologists, enabling us to demonstrate that Dr.Aid's compliance checking spans multiple user actions, potentially conducted by different users, in different organisations with arbitrary time separation.

For both use cases, the provenance traces are obtained from SPARQL endpoints served with Apache Jena Fuseki 3.17[12]. We present each of the two applications in a subsection, covering their relevant features and results, and conclude with a summary afterwards.

*4.2.1 Cyclone tracking.* The cyclone tracking workflow is used to estimate the distribution of tracks of cyclones as a consequence of climate change. It can also track high-pressures and mid-altitude weather systems. Its core component, implemented in Fortran, uses the algorithm and methodology proposed by Sinclair [69]. The workflow is coded in CWL using parallelization (the *scattering* functionality of CWL), and is described in more details in Section 3.2 of [23]. Its provenance is delivered compliant with the CWLProv schema. The workflow is originally a pipeline-style workflow, where the processes are connected one by one. With the parallelization, it becomes three parallel streams merged at the last process. The data used by the original workflow are all obtained from CMIP6[13] whose data-governance rules are presented in [3]. The encoding and discussion are presented in Appendix D, and summarized in Table 2.

Figure 6 shows the identified data-flow graph and activated obligations, by running Dr.Aid on this provenance graph. The extracted data-flow graph (i.e. the top part) corresponds correctly with the original definition of the workflow, with the parallelization described above; its processes have the correct information for each relevant field (e.g. process type, number of input/output ports,

---

[12]Apache Jena Fuseki: https://jena.apache.org/documentation/fuseki2/

[13]CMIP6 website:https://pcmdi.llnl.gov/CMIP6/

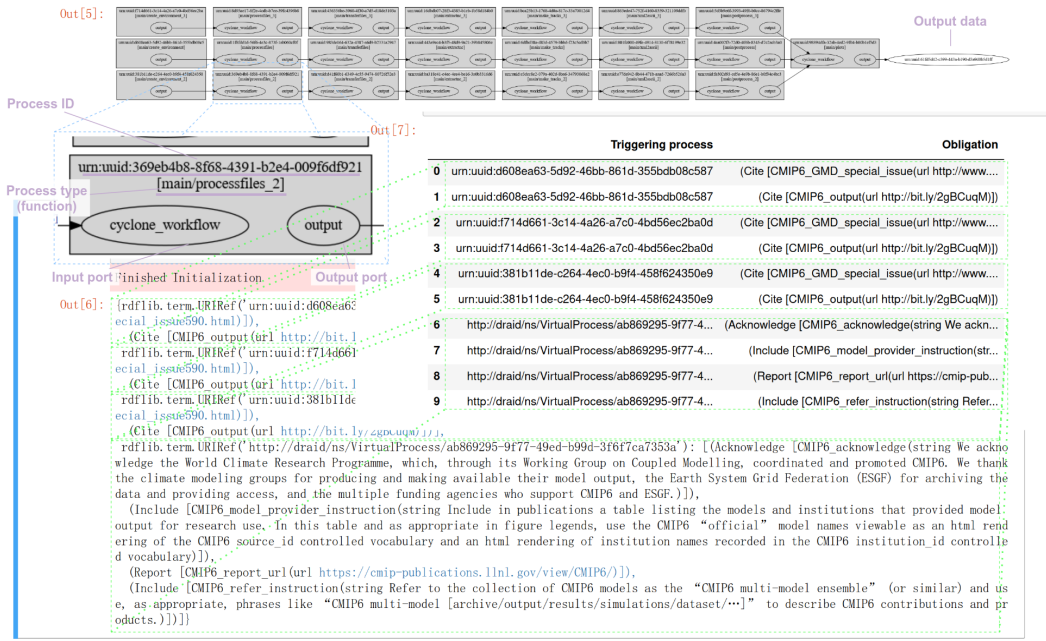

Fig. 6. Visualization of and identified obligations from reasoning about the data-flow graph of the cyclone-tracking workflow.

The top diagram is the visualization of the data-flow graph (in our intermediate representation) extracted from the provenance, with intermediate data objects hidden; it receives some extra annotations (shown magnified) to clarify important aspects; the printed Python dictionary at the bottom is the identified activated obligations; the table to the right is the information stored in the obligation database, corresponding to the dictionary result at the bottom.

name of the ports, etc), as shown and labelled in the magnified process. This means our system is able to correctly extract the data flow information from the provenance data of the cyclone tracking case. In particular, because the process and information are extracted, the original processes in the workflow can be linked. Therefore, the rules associated with them (and data) can also be obtained so long as standard protocols exist. In our evaluation, this is mocked up with an internal database. Readers may observe some seemingly duplicated obligations, activated by different processes, which are expected because of the semantics: the data are used in parallel, and therefore each trace creates one activation following the definition in the data rules (more precisely, the activation condition stage = import). It is an open question whether to keep them, deduplicate them, or to provide another mechanism for specifying them in the rules, which is beyond this paper. As a quick solution, in a deployed system, a user-interface may present the logically distinct obligations. Apart from them, we introduced a virtual process `publish` at the end of the data flow, to represent the human action after producing the output data. It is recognized by the system and activated more obligations, for those with activation condition action = publish. In addition to the identified activated obligations, the reasoning result also contains the derived data rules for each output data. That is shown in Figure 8 in Appendix C.1. (Figures in their original sizes are included in the supplementary material.)

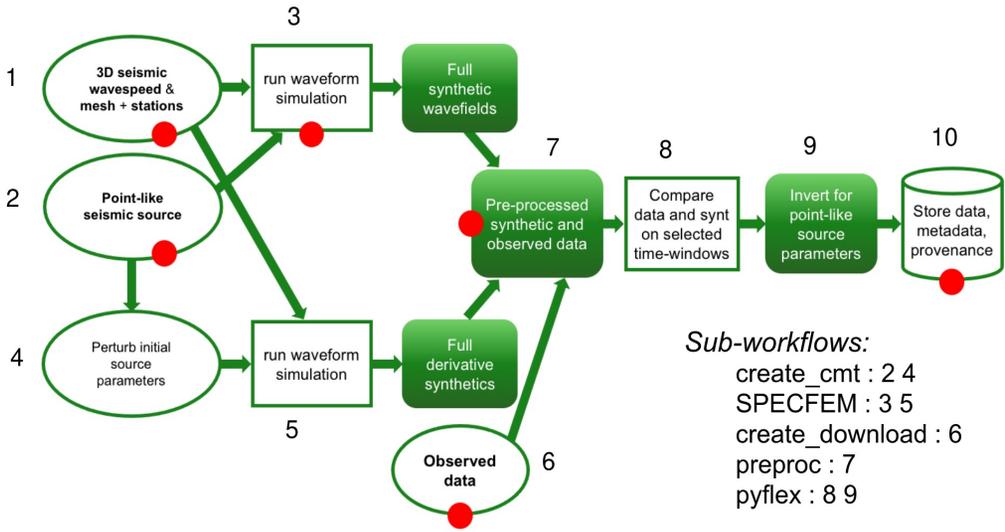

Fig. 7. The conceptual structure of the MT3D workflow, and the sub-workflows with their corresponding steps. Color and shape are unrelated to its usage in our evaluation; figure is based on Figure 3.2 from [23].

*4.2.2   MT3D.* Moment Tensor in 3D (MT3D) is a seismology use case used to study wave propagation and hazard assessment through characterizing the earthquake properties, including the source parameters and their uncertainties. The Earth is represented in a 3D spectral-element model (SEM) of wave speeds. Unlike cyclone tracking, the MT3D workflow is not a single workflow, but comprises several sub-workflows which are executed consecutively, with independent provenance traces that need to be correlated (see Figure 7). Most of the sub-workflows use `dispel4py`, and provenance traces are in S-Prov schema; while the waveform simulation code, SPECFEM3D [64] is driven using CWL. The evaluation performs reasoning on these traces one by one. MT3D has multiple input data for different purposes:

- SEM mesh modelling the Earth's structure;
- Initial parameters identifying the earthquake source;
- The observed earthquake data from seismometers to correlate with model output to estimate errors and iteratively improve the source model.

They can come from different sources, e.g. EIDA[14], INGV[15], Global CMT Catalogue[16], etc. In our experiment, the mesh and wavespeed profiles for the SEM modelling are obtained from personal communications, the parameters of earthquake source are from INGV, and observed earthquake data are from EIDA. The policy for the personal communication, as we obtained from the workflow's author, was a requirement to properly acknowledging the data provider. The properties and encoding of the publicly available policies are summarized in Table 2 in Section 4.4; the policy for the personal communication is also encoded but not included in that table. All encodings and their justifications are presented in Appendix E. As a summary, the model was successful in encoding all the actioning rules.

---

The reasoning results for MT3D are also as expected, attached in Appendix C.2 (also in the supplementary material). Each figure resembles the reasoning result of one sub-workflow (one provenance graph), listed in the order they were executed. These (sub-)workflows are constructed not by explicitly using input data (not passed in directly through input ports), but by internally hard-coding them in the workflow. Thus, we used *virtual input ports* to inject the input data rules (the egg-shaped ports with dashed lines). The reasoning results from the previous traces are then used by the subsequent traces where the data are correlated (e.g. the combined data rules from `SPECFEM` is used as one input to `preproc`, as shown in the output data rules in Figure 10 and the input `import_synt` in Figure 12). The prevailing data rules associated with derived data are retained until the end (i.e. the `pyflex` step, shown in Figure 13). Because some data rules use the trigger of action = publish, these obligations get triggered at the `publish` (virtual) process, which is attached to the last sub-workflow `pyflex` (Figure 13) because it produces the final results; it is not triggered in previous sub-workflows, because they do not have a process with type *publish*. This demonstrates the system can trigger rules at the correct point based on the specification. As above, the virtual processes resembles human actions that the human performs afterwards, which can also be replaced with a computational process of type `publish` if such a process is automated, and Dr.Aid will recognize it too. In addition, as shown in Figure 12, the left-most process explicitly uses flow rules to regulate the flow of (data) rules from the expected input ports to the corresponding output ports. That reflects the underlying data flow within that process, where rules flow with the data, demonstrating the necessity and usefulness of the *flow rule* mechanism. In total, this use case shows that Dr.Aid is able to correctly retain and reason about data rules for multiple separately executed data flow traces (even when they were from different workflow systems). It also provides an example to demonstrate the necessity and usefulness of flow rules.

*4.2.3 Summary.* As a summary, for both use cases, involving six provenance graphs with CWLProv and S-Prov schemas, our framework is able to encode all actioning rules for the data-use policies, correctly extract the necessary information from the provenance traces, and correctly perform the reasoning of the flow rules and data rules, including executing flow rules, associating the expected data rules with data products, deriving subsequent data rules and triggering expected obligations at the correct time. We conclude that our implementation addresses effectively all five rule-handling issues we have identified and has the potential to do this in a wide range of applications.

As discussed in the begging of Section 3, we hold the assumption that, in the foreseeable future, researchers will be able to fetch data and compose workflows using processes from repositories and catalogues, and those data and processes will have rules associated with them (e.g. by the data provider and process authors). The researchers can proceed with their research without worrying about specifying policies for most cases. When they develop new tools or processes, they or their data governors may optionally provide the flow rules for those processes. Given the fact that the language for flow rules is simple, and the authors do not need to know details about the data rules, it should be easy to master and use it. An empirical study for confirming (or refuting) this may need to be performed as future work.

## 4.3 Comparison with other frameworks

In this part, we provide a comparison of Dr.Aid against other frameworks, to discuss their features in more depth (our observations in Section 2 are not repeated here). The example scenario used in Sections 1 and 3 is reused here, and we focus on the use of UA, because it is demonstrative enough. We examine two typical systems: CamFlow and Thoth. The reason we chose them is because of one critical requirement for modelling the example scenario: to automatically change the policies for the output data (*dynamic application*).

The encoding using Dr.Aid has been introduced in Section 3.2. The encoding using other frameworks will be introduced below, and finally a discussion will be presented. Because no other frameworks support all the required features, we will extrapolate from their literature as far as we can to encode unsupported requirements.

We refer to the policies associated with data as *data rules* and the policies associated with processes as *process rules*.

*4.3.1 CamFlow.* The data policy for CamFlow contains two sets of tags, one called security label S and the other called integrity label I. We model the DoB field as a security tag $\mathrm{DoB}$; it does not support modelling the *requirement* of reporting usage nor the acknowledgment requirement (both are *obligations*), so they are dropped. Therefore, the data policy (*security context* of the data entity) for data D is $\{S(D) = \{DoB\}, I(D) = \emptyset\}$. The processes using the data should have their relevant process rule (*security context* of the process), and all of them (denoted as Proc) should have the minimum as the data: $\{S(Proc) = \{DoB\}, I(Proc) = \emptyset\}$. For the process by user A ($Proc_{UA}$) that converts DoB to YroB (i.e. produces data DC), it additionally needs this *label change* specification $S(Proc_{UA}) = \{DoB\} \rightsquigarrow S(Proc'_{UA}) = \{YroB\}$ and to be granted the corresponding privileges $YroB \in P_S^+(Proc_{UA})$, $DoB \in P_S^-(Proc_{UA})$. CamFlow does not provide a way to specify rules for multiple outputs from one process, so UA needs to create another process that produces data DB which has label change specification $S(Proc_{UA}) = \{DoB\} \rightsquigarrow S(Proc'_{UA}) = \emptyset$ and privilege $DoB \in P_S^-(Proc_{UA})$. Similarly, the process by user UX or UY needs to be specified accordingly.

*Discussion.* CamFlow can represent the policy change in a decentralized way regarding the content change, though it still requires the system admin to assign trust to the process before it can acquire the privilege; it can not model the reporting requirement, and also can not model the acknowledgment requirement. A workaround for the reporting requirement is to use the *integrity* set: assign the data with the integrity tag $I(D) = \{acknowledge\}$; all users agree to properly acknowledge the data assign their processes with $I(Proc) = \{acknowledge\}$, But this requires external checks of the proper permission, and the semantics is not represented in the policy too – this is environment information outside the encoded policy.

*4.3.2 Thoth.* The data policy for Thoth contains two layers, three sets of policies: *read* and *update* in layer 1, and *declassify* in layer 2. Because of its model, the data provider DP needs to specify all policies in the very beginning when distributing the data. Therefore, we can assume DP has a metadata file users containing the list of users (by their session keys) that said they agree to report their use of the data back to DP. Then we use the *read* policy to verify if the user is within this list. Thoth does not support checking the process information, so we store the recognized derivation in a metadata file derive, with each tuple of a recognized derivation behaviour, in the form of removeDoB(KEY) and DoBToYroB(KEY) where KEY is the session key. Then we can use the *declassify* policy to specify what policies the derived data should have. Thus, we have this policy set:

| | |
|---|---|
| **read** | : $\neg sKeyIs(k_x) \wedge$ ("users", off) says $k_x$ |
| **declassify** | : $\neg isAsRestrictive(read, this.read)$ until |
| | $sKeyIs(k_x) \wedge ($ |
| | $(("derive", off_1)$ says $removeDoB(k_x) \wedge POLICY\_NO\_DOB) \vee$ |
| | $(("derive", off_2)$ says $DoBToYroB(k_x) \wedge POLICY\_YROB))$ |

where the *read* policy says the current session key is $k_x$ ($sKeyIs(k_x)$) and the key $k_x$ is in the file users (at an offset *off*) (("users", off) says $k_x$); the *declassify* policy says the file and its derivations has the same *read* constraint ($\mathsf{isAsRestrictive(read, this.read)}$), and some of its derived files (i.e. DB) no longer bind to this requirement when the derived file is produced in a session with key $k_x$ and removeDoB($k_x$) is within the file derive ($\mathsf{sKeyIs}(k_x) \land ((\text{“derive”}, \mathsf{off}_1) \text{ says } \mathsf{removeDoB}(k_x))$), and the new file is bound to the new policy POLICY_NO_DOB (which is a placeholder macro for DP to specify the policy for the derived file without the field DoB); similarly, after going through a session matching DoBToYroB($k_x$), the policy for the derived data becomes POLICY_YROB. We do not specify the details of POLICY_NO_DOB and POLICY_YROB because that is much longer, and the example given is sufficient to demonstrate the encoding.

*Discussion.* For Thoth, policies for derived data need to be specified by the data provider DP, which is a large drawback for the federated context, as it is impossible to predict the users and how they are going to use the data exactly. We modelled the acknowledgement and protecting requirement as the user registration information in a metadata file, which does not faithfully resemble the original policy, but is the best effort we can think of. We assumed a known list of processes (which needs constant updates), and thus established the *declassify* rule to pre-specify what derived rules would apply for each type of derived data.

*4.3.3 Summary.* Our finding from these encodings is that, based on their publications, neither CamFlow nor Thoth can model the data-use policies in our example scenario. Both CamFlow and Thoth can represent the policy change with processing (*dynamic application*), though Thoth cannot recognize processes and a non-perfect workaround is used; both of them can differentiate the policies between directly removing DoB and gradually replacing DoB with YroB (and finally removing YroB). However, it is not possible to model the obligations using these frameworks. Neither of them supports modelling (the policy of) multiple outputs from a single process, nor do they provide constructs to deal with multiple inputs for a single process. In addition, Thoth requires the data provider to predict all potential uses of data and specify the policy in advance, not suitable for federated contexts. In short, these frameworks cannot fully encode the policies in the example scenario we posed, while Dr.Aid is able to encode them.

## 4.4 Encoding real-world public data-use policies

To evaluate our model's wider applicability, we examine it using published data-use policies. The main focus is the capability of our model, that is to what extent can our model represent the rules of those policies. We evaluate that by encoding them in our formal representation. We then compare our formalization with the original policy.

*4.4.1 Evaluation design.* We first identify and collect published data-use policies from a range of data providers and archival services typical of the resources used by practitioners working on data-driven research. These *policies* are publicly available, and the data they govern are often also publicly available, though not always (e.g. MIMIC [13]). Then, we follow our standard procedure to convert from the natural-language policies to our formal representation. We record the results, and provide different metrics supporting comparison; Table 2 summarizes these. We then present our interpretation of this evaluation.

*Policy origin.* The policies were collected from publicly available sources. These were found by asking the research scientists what dataset they would use and tracking back to find the relevant data-use policies. We also navigated to the related datasets that these services referenced (e.g. by following the link on their website). Another source was searching for datasets and policies on the Internet. (Contrary to our intuition, the latter method did not produce many useful results.) It is also worth noting that the collected target policies are the data-use policies for the *data users* to comply with. Such policies may have a legal formal backing, but that formality is not our target.

Table 2. Result summary for 15 published data-use policies showing the coverage of the formalization.

| Policy source | | number of | | | | rule coverage | |
|---|---|---|---|---|---|---|---|
| | sentences | rules | actioning | implied | encoded | actioning | total |
| CMIP6[3] | 35 | 9 | 8 | 1 | 7 | 100% | 89% |
| EIDA[8] | 20 | 5 | 3 | 0 | 3 | 100% | 60% |
| INGV[7] | 2 | 2 | 2 | 0 | 2 | 100% | 100% |
| CC-BY[5] | 12 | 6 | 5 | 2 | 3 | 100% | 83% |
| CMT Catalogue[9] | 15 | 4 | 4 | 0 | 4 | 100% | 100% |
| CORDEX[4] | 22 | 9 | 6 | 0 | 5 | 83% | 55% |
| ISMD[12] | 2 | 1 | 1 | 0 | 1 | 100% | 100% |
| RCMT[10] | 14 | 3 | 3 | 2 | 1 | 100% | 100% |
| MIMIC[13] | 17 | 4 | 4 | 0 | 4 | 100% | 100% |
| CPRD[6] | 21 | 7 | 6 | 0 | 2 | 33% | 29% |
| PIMA[15] | 2 | 1 | 1 | 0 | 1 | 100% | 100% |
| ISC[2] | 21 | 7 | 7 | 0 | 7 | 100% | 100% |
| IRIS[11] | 28 | 10 | 10 | 0 | 10 | 100% | 100% |
| OGL[14] | 30 | 7 | 4 | 3 | 1 | 100% | 57% |
| World Bank[16] | 40 | 12 | 7 | 2 | 3 | 71% | 42% |
| **Total** | 281 | 87 | 71 | 10 | 54 | 90% | 74% |

*Figures and Metrics.* Based on the information, the two main indices we evaluate on are:

(1) actioning rule coverage: the proportion of actioning rules in the policy that are encoded;
(2) total rule coverage: the proportion for all rules, not limited to actioning rules, of the policy that are encoded.

And they are defined as:

$$\text{actioning rule coverage:} \quad C_{act} = \frac{N_{enc} + N_{imp}}{N_{act}}$$

$$\text{total rule coverage:} \quad C_{tot} = \frac{N_{enc} + N_{imp}}{N_{rule}}$$

where $N_{enc}$ is the number of rules encoded, $N_{imp}$ is the number of rules implied, $N_{act}$ is the number of actioning rules, and $N_{tot}$ is the total number of rules.

*4.4.2 Result and discussion.* The results are summarized in Table 2, and the encodings are available in Appendix F (except Appendices D for cyclone tracking, E for MT3D). As can be seen, our model is able to represent a substantial proportion of the actioning rules, with $\sum C_{act} \approx 90\%$. This demonstrates that our model is valid as a step towards characterizing real-world data-use policies. It does not reach 100% because of the limitation of the current semantics – only *obligations* are supported, while the real-world policies contain a small number of other rule types, such as prohibitions. It has a lower rate in representing non-actioning rules, with $\sum C_{tot} \approx 74\%$, because of the emphasis of the framework. The primary design goal was to help users comply with policies and share derived data respecting those policies, so the contextual and disclaimer information/rules are not included in the current model. This can be improved by extending the semantics to include such rules; planned in our future work. On the other hand, this ratio of $\sum C_{tot} \approx 74\%$ also shows our model has captured the most important aspects of data-use policies (in the 15 collected policy sets). Because of the extensibility of the model, we have a fair confidence that it can be extended in the future to improve the coverage. Digging into the details, we have some additional findings discussed below.

*Acknowledgement.* All policies present the need for proper acknowledgement of the data author or provider (and/or dataset, data service) in subsequent publications, and some of them have multiple acknowledgement requirements. Our model is able to encode such requirements easily as obligations. The activation conditions are well represented in our model, as it models user actions as virtual processes and treats them uniformly with computational processes, demonstrate in Section 4.2. However, not all of the original policies have a clear specification of the triggering time when such requirements are activated, which may leave ambiguity for the data users. For example, in CMIP6, the condition is explicitly specified as "in publication" (truncated with "..."):

> Include in publications an acknowledgment with language similar to: "We acknowledge the World Climate Research Programme..."

which is encoded as (extracted from Appendix D):

```
Obligation( Acknowledge CMIP6_acknowledge , [ ], action = publish )
Attribute( CMIP6_acknowledge, "We acknowledge the World Climate Research Programme..." )
```

While in ISMD policy, the condition is referred to as "properly":

> Permission to use, copy or reproduce parts of the ISMD-DB is granted provided that ISMD v2.1 is properly referenced as: Marco Massa...

Based on the context, we can infer that it most likely requires users to cite it in their publications, so we encoded it as (extracted from Appendix F.4):

```
Attribute( ISMD_ack, str "Marco Massa..." )
Obligation( Acknowledge ISMD_ack, [ ], action = publish )
```

*Nested policies.* Many policies have nested policies which refer to another policy in addition to the rules stated directly. This enhances the usefulness of the automated framework to facilitate compliance, because such nested policies can be automatically included. Whether a policy contains nested policies is mentioned in its relevant part in the appendix. One example is the EIDA policy, which says:

> Some of the data sets distributed by EIDA have DOI's (Digital Object Identifier) associated with their seismic networks according to a standard procedure recently approved by the FDSN.

where the "seismic networks" is a hyperlink to another webpage[17] that contains the list of all sources, where each source is another hyperlink to its webpage describing its own citation requirement (in particular, its Digital Object Identifier (DOI)) and other information. When dealing with it manually, the researcher needs to jump between several hyperlinks to obtain the full list of policies; when dealing with it automatically (e.g. using our rule language and reasoning system), the detailed different policies can be directly associated with different data, and dealt with by the system. In our example, the MT3D use case uses the AC network from EIDA, and therefore our encoded policy is (extracted from E; truncated here for length):

```
Obligation( Cite AC_network , [ ], action = publish )
Attribute( AC_network, string "Institute Of Geosciences, Energy, Water And Environment..." )
Obligation( Acknowledge ORFEUS_EIDA , [ ], action = publish )
Attribute( ORFEUS_EIDA, string "We acknowledge ORFEUS and EIDA ...." )
```

---

[17]EIDA Networks: http://www.orfeus-eu.org/data/eida/networks/

*Types of rules successfully modelled.* In addition to that, our model can represent actioning requirements such as limiting the purpose of use, users permitted to use, and actions about derived data. They cover the majority of the actioning policies we reviewed. In addition, with flow rules, our model is able to specify contextual constraints (which are to be removed with flow rules using delete() in appropriate processes) and changing contextual information for obligations (to be changed by flow rules using edit()).

*Use of derived data.* Most policies don't explicitly specify the extent to which they apply to derived data. Only CC-BY, OGL and World Bank (and policies that include them as nested policies) explicitly specify that they allow the user to redistribute derived data. But considering the context, all data providers do not prohibit users from distributing derived data, except MIMIC and CPRD which both specify rules for medical data. In fact, the data providers for MIMIC and CPRD also do not object to users publishing results using their data, because they have the acknowledgement requirements in their policies. Because the main reason for MIMIC not being publicly available is *"...database, although de-identified, still contains detailed information regarding the clinical care of patients, so must be treated with appropriate care and respect"*. We suspect that the reason for not directly allowing derived data to be shared is due to concerns over revealing sensitive information. This links to the example use case we illustrated, and our framework provides a promising direction.

*Data merging and CPRD.* Our model accommodates data merging as a *by-default* permitted action. This is invariably true for data policies because of their generic role supporting any research or enquiry. It is not true for the CPRD policy (a medical data policy), which has more specific requirements imposed on users of its data *subset*. Two more rules can be partially modelled if we use *ad hoc* methods, raising the coverage to 66% and 57% for that data provider (details in Appendix F.7.1).

*Reflective actions.* Another drawback is that our model is unable to represent the commitments that users are *required to make* (and initiated by the data providers, in contract to obligations initiated by the data users), such as "to provide information on how they used data when required" (e.g. CORDEX, CPRD). This is a potentially useful and straightforward extension in the future, by including an action initiator in the obligation.

*Compression.* As can be observed from the table, the total number of sentences in the original natural-language policies is much larger than the number of rules $\frac{87}{281} \approx 31\%$ (or $\frac{71}{281} \approx 25\%$ if considering only the actioning rules). This shows that information density is not very high in the natural-language policies, due to various reasons, e.g. the necessity to clarify terms, the inclusion of contextual information, duplicated statements, etc. In principle, some of these information can be defined once and shared across policies[18], but the practice does not follow that. Therefore, even if we just consider the compression of policies, it is already a sensible approach to model the rules using a formal language.

### 4.4.3 Summary.
This part evaluated the capability of our language model used in Dr.Aid using real-life data-use policies from different sources, finding that it has a high capacity to encode the policies we found ($\sum C_{act} \approx 90\%$ against actioning rules, and $\sum C_{tot} \approx 74\%$ against all rules). We also identified and discussed some additional findings by reading and encoding the policies, justified the usefulness of encoding the natural-language policies using the formal model and some design choices of our language.

---

[18]E.g. clarification of terms can be met by using URIs or terms in agreed ontologies behind the scenes in future systems, A service can be consulted to find or obtain explanations or definitions of such terms.

## 4.5 Evaluation conclusion

In this section, we provided three evaluations demonstrating three aspects of our framework: its ability to be used in real-world research practice, its advantage against other frameworks, and its capability to encode real-world data-use policies. We conclude that the Dr.Aid framework has a better model for real-world data-use practice and data-use policies. Of course, our current research has limitations, which are discussed in the next section.

## 5 LIMITATIONS AND FUTURE WORK

We consider the Dr.Aid framework a significant step forward, but there are still many open questions, including those exposed by our research. In this section, we present the limitations we are aware of, and our future work.

*User study.* In this paper, we reported an evaluation of the Dr.Aid framework with objective aspects, which are enough to demonstrate its feasibility and the *potential* benefits of our framework. But a more complete evaluation would include HCI aspects, including usability, learnability, acceptability and utility, the language's understandability and perceptions of trustworthiness and ptoductivity for a representative range of potential stakeholders. This deeper study requires establishing or simulating more of the envisaged future context so that realistic embedding in working practices are emulated. That must lie in our future work program due to time and resource limitations.

The current pilot user study was conducted with four research scientists in diverse domains. It broadened our view of the target context, identified real-life but often unencoded data-use policies, and strengthened the evidence motivating and shaping our framework. We are starting to perform a larger study investigating the usability, utility and understandability of the system and language for more of the roles involved in data-intensive research collaborations. The intended subjects will be scientific workers whose work involves data processing, data handling and curation, method development and evidence production.

*Better policy conversion and association.* In our current research, the data-use policies are converted by hand to the formal encoding. This is a must compromise due to the fact that all such policies are written in natural languages. We consider it viable for the evaluation point of view, because one policy set is usually established for collections of data. But in a prospected future, the data prociders should provide the encoded policies together with the natural-language policies, and a standard mechanism (protocol) should be established to fetch the policies. Natural Language Processing (NLP) technologies may be developed to perform automatic conversion, with or without human verification afterwards.

*Collaboration to enable adoption.* We, and others working in this area of research, envisage a future where data owners and creators can formulate rules that balance productivity with necessary restrictions as precisely as they wish and then be confident that all future use of that data, software or tools will comply with those rules. At the same time, individuals, groups and organisations want to form alliances rapidly in response to new requirements and opportunities. They then want to collaborate sustainably while their federation and their working context evolves to meet needs and to exploit the latest advances. To achieve such a future we need to collaborate with other researchers to develop the understanding, standards and protocols to make that possible. We believe that this should be contemporaneously addressed while systems like Dr,Aid are developed to explore, pioneer and support that future collaboration environment.

*Link with DIFC.* As discussed in Section 2.2, the language model for Dr.Aid is related to DIFC, but takes a different direction, resulting in a more extensible and semantically rich language. Using this extensibility, we plan to establish a more formal link with DIFC. We also intend to draw on, and if possible interact with, other research that is clarifying concepts, developing ontologies and investigating languages that describe data rules and their application e.g. [28].

*Language extension.* The language currently only supports obligations as the actioning construct, which shows the advantage of this language model compared to other approaches. But the current model lacks the explicit semantics for prohibitions/permissions, which is an often-found construct in related research. This can be extended in a way similar to the *pre-obligation* in some other research [30, 61] (they would refer to the obligations in our research as *post-obligations* or *ongoing-obligations*). Apart from that, extension can also be made on the activation conditions to support complex conditions. More expressive logic constructs may be needed, which may require us to adopt additional logical foundations.

*Logic deduction and optimization.* We have used situation calculus as the formal foundation for our reasoning mechanism, but did not investigate its potential extensions. For large workflows, the whole-graph reasoning time grows rapidly. Further optimization can be done (e.g. [31, 70]), such that it would be possible to recursively deduce the "flow rules" of the whole *workflow* graph from the flow rules of individual *processes*. That would allow the whole workflow to be treated as a single process when users are not concerned about associating rules with intermediate results; this also enables automatic deduction of flow rules from code level.

## 6 CONCLUSION

In this paper, we identified and addressed an important and urgent need to supply automation to help workers comply with data-use rules, particularly when they collaborate over long periods and in geographically distributed loosely coupled federations. This computer-supported approach will also help practitioners in simpler contexts. We have shown that the data-rules are widespread and have observed that there is a severe shortage of tools to help users find the relevant rules and comply with them at the critical moments. We clarify the requirement by identifying five important objectives for enabling data-rule compliance in federated contexts. Drawing on relevant contemporary research we opened up a general approach by prototyping a framework, Dr.Aid, which successfully addresses all five objectives. We demonstrated this success using two real-world scientific workflows from meteorology and computational seismology. We also assessed our coverage by encoding the rules published by 15 data repositories. This revealed some limitations and motivated our future work.

We believe this is a major step towards a future where all those involved in data use are supported by a framework inspired by Dr.Aid, covering virtually all data-rules and capturing information from the majority of tools and processes so that the framework can be widely deployed. Humans still take responsibility for formulating rules, but with the improved precision and compliance, rules will become more subtle. Data analysts will be reminded of their obligations as they produce results and as data is passed between them. They retain their autonomy, when they want to they selectively review the unfilled obligations the system has collected for them, drill into details and decide which ones they should deal with. Workflow and software developers understand how their products propagate rules, and will specify when rules can be relaxed as a result of processing or when new rules should be added. Administrators and managers can review obligations. Governance can focus on where rules need revision. This depends on two crucial advances: (1) a formal notation for rules that has a form that users understand and can use, and a form that reasoners can understand and use; and (2) a reasoning system that is coupled to the data handling and processing systems in use that delivers relevant information tuned to each role in the data-sharing community.

## ACKNOWLEDGEMENTS

EU project DARE – DARE: Delivering Agile Research Excellence on European e-Infrastructures (777413) – provided partial support for this research. In particular, we thank the DARE researchers who helped us conduct the system evaluation, namely, Christian Pagé and Alessandro Spinuso, especially regarding workflows and provenance traces. Several DARE colleagues offered their opinions views on and experience of rule encoding.

## SUPPLEMENTARY MATERIALS

This paper has supplementary materials, containing the figures for reasoning results in their original size. Additional materials (esp. provenance used in evaluation) can be found using the DOI `10.5281/zenodo.5227391`[19]. The source code can also be found from the author's repository[20].

---

[19]Additional materials: https://doi.org/10.5281/zenodo.5227391
[20]Source code: https://github.com/renyuneyun/dr.aid/tree/cscw2021

# Appendices

## A    ACTIVATION CONDITION SLOTS

The summary of slots in activation conditions is presented in Table 3. The value can be any value literal or a special constant ∗ representing *any* value.

Table 3. Slots of activation conditions

| Slot | From | Meaning |
|------|------|---------|
| action | Provenance | The process *type* |
| stage | Framework | The processing stage that this rule is involved |
| purpose | User specification | The purpose of this workflow execution |
| user | Provenance | The user identifier, retrieved from provenance |
| startTime | Provenance | The date and time of execution |
| processId | Provenance | The ID of the process |

The available values for the "stage" slot are "start-of-workflow" (start of the workflow), "end-of-workflow" (when the workflow finishes)) and "import" (when the rule is imported to the execution for the first time).

## B  AXIOMS FOR THE SITUATION CALCULUS FORMALIZATION

All the *fluents* are listed here:

$$
\begin{aligned}
&\text{Attr}(n, t, v, h, s)\\
&\text{PropAttr}(n, t, v, h, s)\\
&\text{Obligation}(ob, h, cond, p_{in}, s)\\
&\text{PropObligation}(ob, h, cond, p_{in}, p_{out}, s)
\end{aligned}
\tag{1}
$$

All the *actions* are:

$$
\begin{aligned}
&\text{pr}(p_{in}, ps_{out})\\
&\text{edit}(\underline{n}, \underline{t}, \underline{v}, t_{new}, v_{new}, \underline{p_{in}}, \underline{p_{out}})\\
&\text{delete}(\underline{n}, \underline{t}, \underline{v}, \underline{p_{in}}, \underline{p_{out}})
\end{aligned}
\tag{2}
$$

where the underscore marks that this argument may be $*$ which denotes *arbitrary*. We require that the original rule can not contain $*$ as its value for these arguments.

The precondition axioms are simply $\top$ (true), because we expect the action be still perform-able but does nothing when the expected conditions do not hold.

$$
\begin{aligned}
&\text{Poss}(\text{pr}(p_{in}, ps_{out}), s) \Leftrightarrow \top\\
&\text{Poss}(\text{edit}(\underline{n}, \underline{t}, \underline{v}, t_{new}, v_{new}, \underline{p_{in}}, \underline{p_{out}}), s) \Leftrightarrow \top\\
&\text{Poss}(\text{delete}(\underline{n}, \underline{t}, \underline{v}, \underline{p_{in}}, \underline{p_{out}}), s) \Leftrightarrow \top
\end{aligned}
\tag{3}
$$

The successor-state axioms are:

$$
\begin{aligned}
&\text{PropAttr}(n, t, v, h = [h_0|[p_{in}, p_{out}]], do(a, s) \Leftrightarrow\\
&\quad \text{PropAttr}(n, t, v, h, s)\\
&\quad \wedge \neg(a = \text{delete}(\underline{n}, \underline{t}, \underline{v}, \underline{p_{in}}, \underline{p_{out}})\\
&\qquad \vee \exists v_2 \neq v.a = \text{edit}(\underline{n}, \underline{t}, \underline{v}, t_2, v_2, \underline{p_{in}}, \underline{p_{out}})\\
&\qquad \vee a = \text{end}(p_{out}))\\
&\quad \vee \text{PropAttr}(n, t_{old}, v_{old}, h, s) \wedge (a = \text{edit}(\underline{n}, \underline{t_{old}}, \underline{v_{old}}, t, v, \underline{p_{in}}, \underline{p_{out}}))\\
&\quad \vee \text{Attr}(n, t, v, h_1 = [h_0|[p_{in}]], s) \wedge (a = \text{pr}(p_{in}, ps_{out})) \wedge p_{out} \in ps_{out}
\end{aligned}
\tag{4}
$$

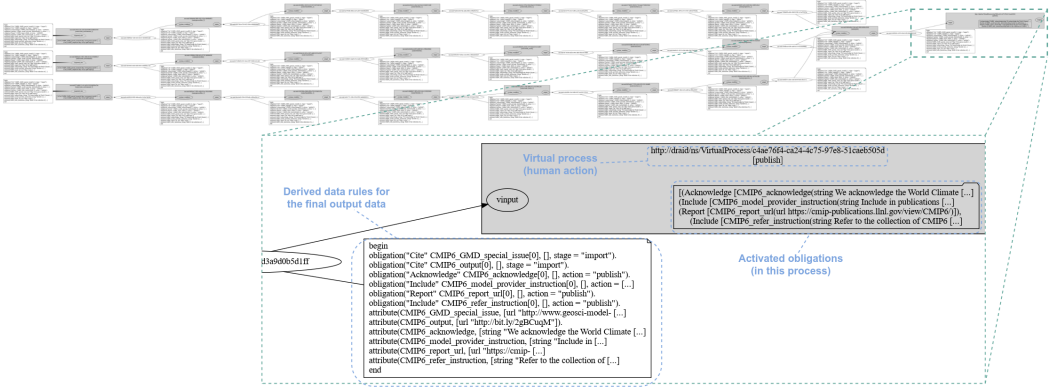

Fig. 8. Derived data rules for cyclone tracking, and the injected virtual process "publish"

$$\text{PropObligation}(ob, h = [h_0|[p_{in}, p_{out}]], cond, p_{in}, p_{out}, do(a, s)) \Leftrightarrow$$
$$\neg(\exists n, t, v, p_{in}, p_{out}.\{\text{PropAttr}(n, t, v, h, s) \wedge a = delete(\underline{n}, \underline{t}, \underline{v}, \underline{p_{in}}, \underline{p_{out}})\}$$
$$\vee a = end(p_{out}))$$
$$\vee \text{Obligation}(ob, h_1 = [h_0|[p_{in}]], cond, p_{in}, s) \wedge a = pr(p_{in}, ps_{out}) \wedge p_{out} \in ps_{out}$$
(5)

$$\text{Attr}(n, t, v, h = [\_|[p]], do(a, s)) \Leftrightarrow$$
$$\text{Attr}(n, t, v, h, s) \wedge \neg\exists psa = pr(p, ps)$$
$$\vee \text{PropAttr}(n, t, v, h, s) \wedge a = end(p)$$
(6)

$$\text{Obligation}(ob, h, cond, p, do(a, s)) \Leftrightarrow$$
$$\text{Obligation}(ob, h, cond, p, s) \wedge \neg\exists ps.a = pr(p, ps)$$
$$\vee \text{PropObligation}(ob, h, cond, p, s) \wedge a = end(p)$$
(7)

The arguments correspond to those explained in Section 3.2, so we omit the explanation for simplicity. In these axioms, we use a notation similar to Prolog's notation of lists when retrieving elements in histories, but we do this in the reversed order to indicate that they are appended, conceptually. Similarly, the = in the head/consequence (e.g. h = [_|[p]]) means expansion (to be used later in the body), rather than assignment.

## C   RESULTS OF FRAMEWORK EVALUATION

### C.1   Cyclone tracking

Here we have the derived data rules for the cyclone tracking workflow in Figure 8.

### C.2   Results for MT3D

See Figure 9, 10, 11, 12, 13 for the reasoning results of each sub-workflow. See Figure 14 for the database containing all activated obligations after running the reasoning for all MT3D sub-workflows.

## D   DATA-GOVERNANCE RULE ENCODING OF CYCLONE TRACKING WORKFLOW

The CMIP6 policy contains multiple rules each may be a link to another nested policy / document.

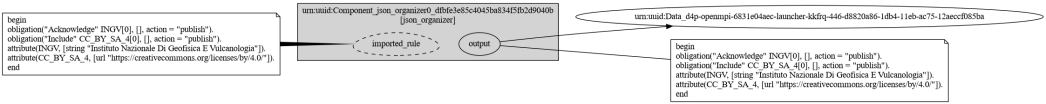

Fig. 9. MT3D reasoning result for create_cmt sub-workflow

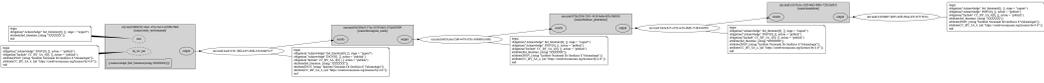

Fig. 10. MT3D reasoning result for SPECFEM sub-workflow

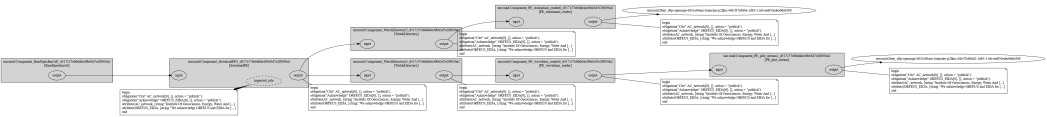

Fig. 11. MT3D reasoning result for download sub-workflow

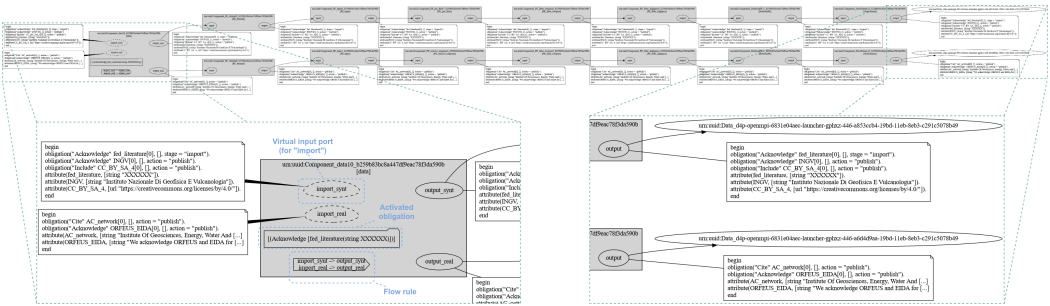

Fig. 12. MT3D reasoning result for preproc sub-workflow

The document pointed to contains duplicated information, e.g. CMIP6 Data Citation Guidelines[21], and we discard them. CMIP6 policy contains contextual information, that specifies how its sub-datasets may have different policies. This is automatically addressed by the framework.

When counting the number of sentences, we consider each acknowledge content as one sentence. The sentences in the CMIP6 Data Citation Guidelines are included, because it is defines additional policies on its own, while the other links do not.

Because most rules do not precisely specify when they should be triggered, we must make assumptions based on the context. We believe most of them should be trigger when the user intends to publish the results, therefore we use action = publish as the activation condition. For demonstration purpose, we model some less-strongly implied rules slightly differently: we say that they will trigger when the data is used by the workflow, i.e. stage = import. This may look the same regarding the eventual result at the first glance, but will constitute to different implications. Our reasoning result demonstrates this: they will be triggered multiple times because of the parallel executions.

Obligation( Cite CMIP6_GMD_special_issue , [ ], stage = import )

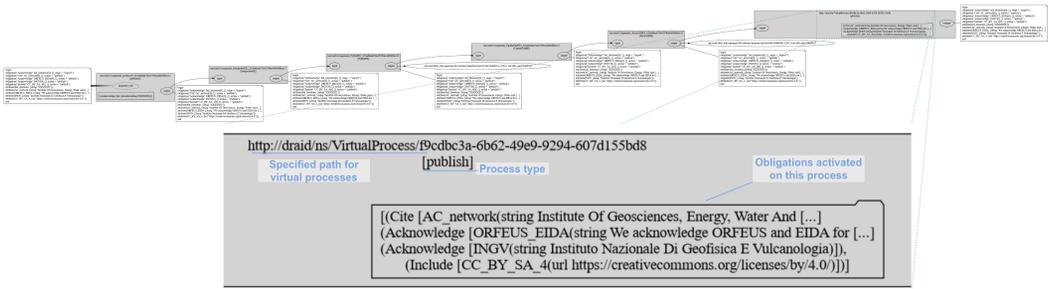

Fig. 13.  MT3D reasoning result for pyflex sub-workflow

```
In [9]: from draid.obligation_store import ObligationStore
        import pandas as pd

        store = ObligationStore(setting.OBLIGATION_DB)
        obs = store.list()
        pd.DataFrame(obs, columns=['Triggering process', 'Obligation'])
```

Out[9]:

|   | Triggering process | Obligation |
|---|---|---|
| 0 | urn:uuid:08fef362-9ae1-47ec-8a14-cb590a706038 | (Acknowledge [fed_literature(string XXXXXX)]) |
| 1 | urn:uuid:Component_data10_b259b83bc8a447df9eac... | (Acknowledge [fed_literature(string XXXXXX)]) |
| 2 | urn:uuid:Component_producer0_b24ed6ada70a4559b... | (Acknowledge [fed_literature(string XXXXXX)]) |
| 3 | http://draid/ns/VirtualProcess/f9cdbc3a-6b62-4... | (Cite [AC_network(string Institute Of Geoscien... |
| 4 | http://draid/ns/VirtualProcess/f9cdbc3a-6b62-4... | (Acknowledge [ORFEUS_EIDA(string We acknowledg... |
| 5 | http://draid/ns/VirtualProcess/f9cdbc3a-6b62-4... | (Acknowledge [INGV(string Instituto Nazionale ... |
| 6 | http://draid/ns/VirtualProcess/f9cdbc3a-6b62-4... | (Include [CC_BY_SA_4(url https://creativecommo... |

Fig. 14.  The stored activation conditions emerged from the MT3D reasoning

Attribute ( CMIP6_GMD_special_issue, url "http://www.geosci−model−dev.net/special_issue590.
    ↪ html" )

Obligation( Cite CMIP6_output , [ ], stage = import )
Attribute( CMIP6_output, url "http://bit.ly/2gBCuqM" )

Obligation( Acknowledge CMIP6_acknowledge , [ ], action = publish )
Attribute( CMIP6_acknowledge, "We acknowledge the World Climate Research Programme,
    ↪ which, through its Working Group on Coupled Modelling, coordinated and promoted
    ↪ CMIP6. We thank the climate modeling groups for producing and making available
    ↪ their model output, the Earth System Grid Federation (ESGF) for archiving the data and
    ↪  providing access, and the multiple funding agencies who support CMIP6 and ESGF." )

Obligation( Include CMIP6_model_provider_instruction , [ ], action = publish )

```
Attribute( CMIP6_model_provider_instruction, string "Include in publications a table listing the
    ↪ models and institutions that provided model output for research use. In this table and as
    ↪  appropriate in figure legends, use the CMIP6 'official' model names viewable as an
    ↪ html rendering of the CMIP6 source_id controlled vocabulary and an html rendering of
    ↪ institution names recorded in the CMIP6 institution_id controlled vocabulary" )

Obligation( Report CMIP6_report_url , [ ], action = publish )
Attribute( CMIP6_report_url , url "https://cmip-publications.llnl.gov/view/CMIP6/" )

Obligation( Include CMIP6_refer_instruction, [ ], action = publish )
Attribute( CMIP6_refer_instruction, string "Refer to the collection of CMIP6 models as the '
    ↪ CMIP6 multi-model ensemble' (or similar) and use, as appropriate, phrases like 'CMIP6
    ↪  multi-model [archive/output/results/simulations/dataset/...]' to describe CMIP6
    ↪ contributions and products." )
```

## E  DATA-GOVERNANCE RULE ENCODING OF MT3D WORKFLOW

Here are the rule encodings involved in examining the MT3D workflow. It is split in three parts, each representing a rule origin.

*For personal communication.* The rule for personal communication is simple, and we directly encode it.

```
Obligation( Acknowledge fed_literature , [ ], stage = import )
Attribute( fed_literature, string "XXXXXX" )
```

*For EIDA.* The EIDA policy contains nested policy, which refers to additional policies in separate webpages. This nesting requires multiple hops to find the required policies. They are all included in our encoding. In the disclaimer, the EIDA policy specifies some additional rules, e.g. not allowing the user to blame the data provider. They are a mix of contextual information and non-actioning rules. When counting the number of sentences, we include also the ones in the nested policies. But we count only the ones about policies itself, not any more. This is an underestimate of the efforts needed when reading the policies manually.

```
Obligation( Cite AC_network , [ ], action = publish )
Attribute( AC_network, string "Institute Of Geosciences, Energy, Water And Environment.
    ↪ (2002). Albanian Seismological Network [Data set]. International Federation of Digital
    ↪ Seismograph Networks. https://doi.org/10.7914/SN/AC" )

Obligation( Acknowledge ORFEUS_EIDA , [ ], action = publish )
Attribute( ORFEUS_EIDA, string "We acknowledge ORFEUS and EIDA for providing the
    ↪ waveform data." )
```

*For INGV.* The data-use policy for INGV contains nested policies of CC-BY. In fact, it says almost nothing more than it is licensed under CC-BY. To avoid duplication, we simply consider CC-BY as a nested policy, and use the Include obligated action to refer to it. When counting the number of sentences, CC-BY is counted as 1 (and 0 implied rules).

```
Obligation( Acknowledge INGV, [ ], action = publish )
Attribute( INGV, string "Instituto Nazionale Di Geofisica E Vulcanologia" )
```

```
Obligation( Include CC_BY_4 , [ ], action = publish )
Attribute( CC_BY_4, url "https://creativecommons.org/licenses/by/4.0/" )
```

## F   ENCODING OF PUBLIC DATA-USE POLICIES

The encodings are presented in each corresponding subsections. Each of them starts with the information and explanation, and then the encoding.

### F.1   CC-BY

CC-BY is a widely known licence for shared work. Its URL is: https://creativecommons.org/licenses/by/4.0/.

When counting the number of sentences, we consider the user-facing version, instead of the legal document oriented for interpretation by lawyers.

CC-BY contains two main types of information: what the user is allowed to do and what the user must comply with when doing so (i.e. requirements). The allowed behaviours are all by-default behaviours in our model; the requirements is written as one sentence but contains three distinct actions – 1) crediting the original material and the author, 2) providing a link to the CC-BY licence, and 3) indicate changes made.

We use a simple encoding first (and use this in the table):

```
Attribute( cc_by, str "https://creativecommons.org/licenses/by"/4.0/ )
Attribute( provider, str "Some–Data–Provider, on Original–"URL )
Obligation( Acknowledge provider, [ cc_by ], action = publish )
Obligation( ProvideLink cc_by, [ ], action = publish )
Obligation( IndicateChanges provider, [ cc_by ], action = publish )
```

In this encoding, the data provider and data url are both specified within the provider attribute. The 1st obligation statement (with Acknowledge) requirement specifies the crediting action; the 2nd obligation statement (with ProvideLink) specifies the link provide action; the 3rd obligation statement (with IndicateChanges) specifies the last action.

User of this data can change the "provider" attribute through flow rules, and therefore allowing further users to compare changes to this output instead of the original data. This is a possible interpretation to the CC-BY's rule of indicating changes.

But there is a drawback that the original data author gets removed too. To solve this, one can define the provider and link as two different attributes. Another drawback is that this encoding pushes all definition jobs to the framework's core language of obligated actions, etc. It doesn't make use of ontologies to specify obligated action classes or attribute names to facilitate such distributed but interoperable context. Therefore, to illustrate how ontologies are used, we assume CC has a separate namespace cc and specifies the classes or names in it. Therefore, we can do an encoding similar to this:

```
Attribute( cc:cc_by, str "https://creativecommons.org/licenses/by"/4.0/ )
Attribute( :provider, str "Some–Data–"Provider )
Attribute( :past_version, url "Original–"URL )
Obligation( cc:Acknowledge :provider :past_version, [ cc:cc_by ], action = publish )
Obligation( :Include cc:cc_by, [ ], action = publish )
Obligation( cc:IndicateChanges :past_version, [ cc:cc_by ], action = publish )
```

In this way, the Dr.Aid framework author is no longer the sole body who can specify the definitions (for action classes, atrribute names, etc). In particular, the definition of cc:Acknowledge is different from the default definition provided by the core language of Dr.Aid. The users are able to change the URL without affecting the original provider too.

Again, they are illustrations of several potential ways to encode the policy. We merely exposed the ambiguities within the original policy by formally modelling them, and provide different solutions to them.

### F.2 Global CMT Catalogue

The page containing the data-use policy is at: https://www.globalcmt.org/CMTcite.html. This policy has nested policies.

The third rule requires proper citation to the exact rules in the website. The idea solution is to use our language to model the rules for each dataset and associate that directly with the data, and thus removing the need to look up. Our language is able to model them, so we assume they are one rule and is properly modelled.

The fourth rule is about the data from old pre-digital collections. It provides three papers, but did not explain how the user should react. We assume this means the user should properly acknowledge either all of them or the used ones. This is within the capability of our model.

There is an option for doing the first two citations or the third or fourth citation (or all of them) in the original rule.

```
Attribute( CMT_meth_app, str "Dziewonski, A. M., T.–A. Chou and J. H. Woodhouse,
    ↪ Determination of earthquake source parameters from waveform data for studies of
    ↪ global and regional seismicity, J. Geophys. Res., 86, 2825–2852, 1981. doi:10.1029/
    ↪ JB086iB04p02825" )
Obligation( Acknowledge CMT_meth_app, [ ], action = publish )

Attribute( CMT_analysis, str "Ekström, G., M. Nettles, and A. M. Dziewonski, The global CMT
    ↪ project 2004–2010: Centroid–moment tensors for 13,017 earthquakes, Phys. Earth
    ↪ Planet. Inter., 200–201, 1–9, 2012. doi:10.1016/j.pepi.2012.04.002" )
Obligation( Acknowledge CMT_analysis, [ ], action = publish )

Attribute( CMT_study_coll, url "http://www.globalcmt.org/Events/" )
Obligation( Cite CMT_study_coll, [ ], action = publish )

Attribute( CMT_analysis, str "
    Ekström, G., and M. Nettles, Calibration of the HGLP seismograph network and centroid–
        ↪ moment tensor analysis of significant earthquakes of 1976, Phys. Earth Planet.
        ↪ Inter., 101, 219–243, 1997. doi:10.1016/S0031–9201(97)00002–2

    Huang, W. C., E. A. Okal, G. Ekström, and M. P. Salganik, Centroid moment tensor solutions
        ↪  for deep earthquakes predating the digital era: The World–Wide Standardized
        ↪ Seismograph Network dataset (1962–1976), Phys. Earth Planet. Inter., 99, 121–129,
        ↪ 1997. doi:10.1016/S0031–9201(96)03177–9

    Chen, P. F., M. Nettles, E. A. Okal, and G. Ekström, Centroid moment tensor solutions for
        ↪ intermediate–depth earthquakes of the WWSSN–HGLP era (1962–1975), Phys.
        ↪ Earth Planet. Inter., 124, 1–7, 2001. doi:10.1016/S0031–9201(00)00220–X
" )
Obligation( Acknowledge CMT_analysis, [ ], action = publish )
```

## F.3 CORDEX

The policy is stated in https://www.hereon.de/imperia/md/assets/clm/cordex_terms_of_use.pdf. This policy has nested policies.

There are different policies for data given to users with different purposes, names research or education or commercial. We model them as three different rules, and different one of them can be attached to the model when distributing the model.

The last rule essentially specifies another acknowledge requirement, but in a less direct way. That requires acknowledging the proper publication associated with the dataset used. This is the direct intention of our framework, so we consider this modelled.

In addition to the normal terms, we added another attribute to represent the scope when the data is still considered as CORDEX (derived) data, and refer to it in all validity bindings. This is optional, and we did this to demonstrate a potential usage of the language and the framework – when a process considers the output is no longer a derivation of CORDEX, it can delete this attribute, and all associated CORDEX obligations are deleted too.

---

Attribute( CORDEX, url, "https://www.hereon.de/imperia/md/assets/clm/cordex_terms_of_use.
    ↪ pdf" )

Obligation( Prohibited, [CORDEX], purpose != research )
Obligation( Prohibited, [CORDEX], purpose != education )
Obligation( Prohibited, [CORDEX], purpose != commercial )

Attribute( CORDEX_ack, str "We acknowledge the World Climate Research Programme's
    ↪ Working Group on Regional Climate, and the Working Group on Coupled Modelling,
    ↪ former coordinating body of CORDEX and responsible panel for CMIP5. We also thank
    ↪ the climate modelling groups (listed in Table XX of this paper) for producing and
    ↪ making available their model output. We also acknowledge the Earth System Grid
    ↪ Federation infrastructure an international effort led by the U.S. Department of Energy's
    ↪ Program for Climate Model Diagnosis and Intercomparison, the European Network for
    ↪ Earth System Modelling and other partners in the Global Organisation for Earth System
    ↪  Science Portals (GO–ESSP)." )
Obligation( Acknowledge CORDEX_ack, [CORDEX], action = publish )

Attribute( CORDEX_doi, str "I understand that Digital Object Identifiers (DOI's used, for
    ↪ example, in journal citations) together with a citation reference will be assigned to
    ↪ some of the CORDEX datasets during the DataCite data publication process, and when
    ↪ available and as appropriate, I will cite CORDEX data by these citation references in my
    ↪  publications. I will consult the CORDEX data website (http://cordex.dmi.dk) to learn
    ↪ how to do this." )
Obligation( Include CORDEX_doi, [CORDEX], action = publish )

---

## F.4 ISMD

---

Attribute( ISMD_ack, str "Marco Massa, Ezio 'DAlema, Sara Lovati, Simona Carannante,
    ↪ Gianlorenzo Franceschina, Paolo Augliera (2016). INGV Strong Motion Data (ISMD) v2
    ↪ .1, Istituto Nazionale di Geofisica e Vulcanologia (INGV). https://doi.org/10.13127/ismd
    ↪ .2.1" )
Obligation( Acknowledge ISMD_ack, [ ], action = publish )

---

## F.5 RCMT

The policy is stated directly on http://rcmt2.bo.ingv.it/, which has nested policies. The data is licensed under CC-BY.

It also synthesizes data from several different sources, each has their own policies with the acknowledgment requirement. We stop here, as this policy did not indicate that the user should also provide acknowledgment to them.

```
Attribute( RCMT_ack, str "Pondrelli, S. (2002). European–Mediterranean Regional Centroid–
    ↪ Moment Tensors Catalog (RCMT) [Data set]. Istituto Nazionale di Geofisica e
    ↪ Vulcanologia (INGV). https://doi.org/10.13127/rcmt/euromed" )
Obligation( Acknowledge RCMT_ack, [ ], action = publish )
Obligation( IndicateChanges, [ ], action = publish )
```

## F.6 MIMIC

The policy is stated in this page https://mimic.physionet.org/about/acknowledgments/, and some additional information are in https://mimic.physionet.org/gettingstarted/access/.

This repository contains rules for two types of assets, the MIMIC data and the MIMIC code. Different rules apply to them.

### F.6.1 Data.

```
Attribute( MIMIC_ack, str "MIMIC–III, a freely accessible critical care database. Johnson AEW,
    ↪ Pollard TJ, Shen L, Lehman LH, Feng M, Ghassemi M, Moody B, Szolovits P, Celi LA,
    ↪ and Mark RG. Scientific Data (2016). DOI: 10.1038/sdata.2016.35. Available at: http://
    ↪ www.nature.com/articles/sdata201635" )
Obligation( Acknowledge MIMIC_ack, [ ], action = publish )

Attribute( MIMIC_data, str "Pollard, T. J. & Johnson, A. E. W. The MIMIC–III Clinical Database
    ↪ http://dx.doi.org/10.13026/C2XW26 (2016)." )
Obligation( Acknowledge MIMIC_data, [ ], action = publish )

Attribute( PhysioNet_ack, str "Physiobank, physiotoolkit, and physionet components of a new
    ↪ research resource for complex physiologic signals. Goldberger AL, Amaral LAN, Glass L
    ↪ , Hausdorff JM, Ivanov P, Mark RG, Mietus JE, Moody GB, Peng C, and Stanley HE.
    ↪ Circulation. 101(23), –pe215e220. 2000." )
Obligation( Acknowledge PhysioNet_ack, [ ], action = publish )
```

### F.6.2 Code.

```
Attribute( MIMIC_code, str "Johnson, Alistair EW, David J. Stone, Leo A. Celi, and Tom J.
    ↪ Pollard. "The MIMIC Code Repository: enabling reproducibility in critical care research."
    ↪ Journal of the American Medical Informatics Association (2017): ocx084." )
Obligation( Acknowledge MIMIC_code, [ ], action = publish )
```

## F.7 CPRD

The post-use policy is stated for each dataset on https://www.cprd.com/DOIs. There are multiple datasets each with their own DOIs. We use one of the real datasets in the example encoding, because the synthetic datasets contains fewer rules.

In addition, accessing their data requires application by going through https://www.cprd.com/data-access where additional policies are stated in the application form.

The special part of it is that the data and results shall be kept confidential and used only by the applicant, which is what the Prohibited obligations state. But this can be lifted under certain conditions, which can be expressed as a process removing the CPRD_controlled attribute (thus removing the bound obligations).

---

Attribute( CPRD_gold_mar, str "Citation: Clinical Practice Research Datalink. (2021). CPRD
↪ GOLD March 2021 (Version 2021.03.001) [Data set]. Clinical Practice Research Datalink.
↪ https://doi.org/10.48329/WH2F"−8168 )
Obligation( Acknowledge CPRD_gold_mar, [ ], action = publish )

Attribute( CPRD_controlled, url "https://www.cprd.com/Data−"access )
Obligation( Prohibited, [CPRD_controlled], action = publish )
Obligation( Prohibited, [CPRD_controlled], user != ""SomeUserId )

---

*F.7.1 Ad hoc modelling for better coverage.* This part describes the *ad hoc* method we mentioned in 4 to increase the encoding coverage. The method is to enforce representing the specified "prohibited" actions as processes, and use flow rules to exploit them. Such actions include:

> Data is not to be used to identify, contact or target patients or general medical practitioners Data is not to be used to study the effectiveness of advertising campaigns or sales forces

They can be modelled as:

---

Attribute( patient, column 3 )
Attribute( medical_practitioner, column 4 )
Obligation( Prohibited, [CPRD_controlled, patient, medical_practitioner], action = identify )

Obligation( Prohibited, [CPRD_controlled], action = forAdvertisingCampaigns )
Obligation( Prohibited, [CPRD_controlled], action = forSalesForces )

---

We call them ad hoc because they only works if the users are under strict constraints with the data providers / governors, so they can agree on a specific way to label the processes (instead of the expected usual way, i.e. to label them with regard to their processing of the data). The processes must be labelled with their specific intention. A possible solution to this is to introduce process type type ontologies with multi-parenting hierarchy.

## F.8  PIMA

This dataset is licensed under CC-0, but proper acknowledgement is encouraged. This page contains relevant information: https://www.kaggle.com/uciml/pima-indians-diabetes-database.

---

Attribute( PIMA_ack, str "Smith, J.W., Everhart, J.E., Dickson, W.C., Knowler, W.C., & Johannes,
↪ R.S. (1988). Using the ADAP learning algorithm to forecast the onset of diabetes
↪ mellitus. In Proceedings of the Symposium on Computer Applications and Medical Care
↪ (pp. 261−−265). IEEE Computer Society Press." )
Obligation( Acknowledge PIMA_ack, [ ], action = publish )

---

## F.9 ISC

There are multiple sub-datasets contained in this data source. The collective policy is accessible through http://www.isc.ac.uk/citations/.

This policy contains nested policies for different sub-items. Each of them has different specific policies, but the general form is to properly acknowledge the dataset and the research work being used. Therefore, we use the first one of them, ISC Bulletin, as the encoding example.

```
Attribute( ISC_product, str "International Seismological Centre (20XX), On−line Bulletin, https://
    ↪ doi.org/10.31905/D808B830" )
Obligation( Acknowledge ISC_product, [ ], action = publish )

Attribute( ISC_art_a, str "Bondár, I. and D.A. Storchak (2011). Improved location procedures at
    ↪ the International Seismological Centre, Geophys. J. Int., 186, 1220−1244, doi: 10.1111/j
    ↪ .1365−246X.2011.05107.x" )
Obligation( Acknowledge ISC_art_a, [ ], action = publish )

Attribute( ISC_art_b1, str "Storchak, D.A., Harris, J., Brown, L., Lieser, K., Shumba, B., Verney, R.,
    ↪ Di Giacomo, D., Korger, E. I. M. (2017). Rebuild of the Bulletin of the International
    ↪ Seismological Centre (ISC), part 1: −19641979. Geosci. Lett. (2017) 4: 32. doi: 10.1186/
    ↪ s40562−017−0098−z" )
Obligation( Acknowledge ISC_art_b1, [ ], action = publish )

Attribute( ISC_art_b2, str "
Storchak, D.A., Harris, J., Brown, L., Lieser, K., Shumba, B., Di Giacomo, D. (2020) Rebuild of the
    ↪ Bulletin of the International Seismological Centre (ISC)—part 2: −19802010. Geosci. Lett.
    ↪ 7: 18, https://doi.org/10.1186/s40562−020−00164−6" )
Obligation( Acknowledge ISC_art_b2, [ ], action = publish )

Attribute( ISC_art_c, str "R J Willemann, D A Storchak (2001). Data Collection at the
    ↪ International Seismological Centre, Seis. Res. Lett., 72,, 440−453, doi: https://doi.org
    ↪ /10.1785/gssrl.72.4.440" )
Obligation( Acknowledge ISC_art_c, [ ], action = publish )

Attribute( ISC_art_d, str "Di Giacomo, D., and D.A. Storchak (2016). A scheme to set preferred
    ↪ magnitudes in the ISC Bulletin, J. Seism., 20(2), 555−567, doi: 10.1007/s10950
    ↪ −015−9543−7" )
Obligation( Acknowledge ISC_art_d, [ ], action = publish )

Attribute( ISC_art_e1, str "Lentas, K., Di Giacomo, D., Harris, J., and Storchak, D. A. (2019). The
    ↪ ISC Bulletin as a comprehensive source of earthquake source mechanisms, Earth Syst.
    ↪ Sci. Data, 11, 565−578, doi: https://doi.org/10.5194/essd−11−565−2019" )
Obligation( Acknowledge ISC_art_e1, [ ], action = publish )

Attribute( ISC_art_e2, str "Lentas, K. (2018). Towards routine determination of focal mechanisms
    ↪ obtained from first motion P−wave arrivals, Geophys. J. Int., 212(3), −16651686. doi:
    ↪ 10.1093/gji/ggx503" )
Obligation( Acknowledge ISC_art_e2, [ ], action = publish )
```

Attribute( ISC_art_f, str "Adams, R.D., Hughes, A.A., and McGregor, D.M. (1982). Analysis
    ↪ procedures at the International Seismological Centre. Phys. Earth Planet. Inter. 30:
    ↪ 85–93, doi: https://doi.org/10.1016/0031–9201(82)90093–0" )
Obligation( Acknowledge ISC_art_f, [ ], action = publish )

## F.10   IRIS

This data source also contains diverse data and therefore diverse rules, stated on https://www.iris.edu/hq/iris_citations. It also has nested policies to FSDN.

This policy set contains different policies for different assets. Most rules are simply requiring the user to properly acknowledge the data being used.

The second rule is about properly acknowledging FDSN object. This is the same as for EIDA data. Therefore, for simplicity, we treat this as one rule in this example, and use the Cite obligated action.

Attribute( IRIS_report, url "https://www.iris.edu/hq/forms/submit_citation" )
Obligation( Report IRIS_report, [ ], action = publish )

Attribute( IRIS_service, str "The facilities of IRIS Data Services, and specifically the IRIS Data
    ↪ Management Center, were used for access to waveforms, related metadata, and/or
    ↪ derived products used in this study. IRIS Data Services are funded through the
    ↪ Seismological Facilities for the Advancement of Geoscience (SAGE) Award of the
    ↪ National Science Foundation under Cooperative Support Agreement EAR–1851048." )
Obligation( Acknowledge IRIS_service, [ ], action = publish )

Attribute( IRIS_FDSN, url "https://www.fdsn.org/networks/citation/" )
Obligation( Cite IRIS_FDSN, [ ], action = publish )

Attribute( IRIS_GSN, str "Global Seismographic Network (GSN) is a cooperative scientific facility
    ↪  operated jointly by the Incorporated Research Institutions for Seismology (IRIS), the
    ↪ United States Geological Survey (USGS), and the Seismological Facilities for the
    ↪ Advancement of Geoscience (SAGE) Award of the National Science Foundation (NSF),
    ↪ under Cooperative Support Agreement EAR–1851048." )
Obligation( Acknowledge IRIS_GSN, [ ], action = publish )

Attribute( IRIS_PASSCAL_Polar, str "Acknowledgment – In any publications or reports resulting
    ↪  from the using IRIS' Polar–specific instruments or support, please include the
    ↪ following statement in the acknowledgment section. You are also encouraged to
    ↪ acknowledge NSF and IRIS in any contacts with the news media or in general articles.\
    ↪ nThe seismic instruments were provided by the Incorporated Research Institutions for
    ↪ Seismology (IRIS) through the PASSCAL Polar Support Services. Data collected will be
    ↪ available through the IRIS Data Management Center. The facilities of the IRIS
    ↪ Consortium are supported by the National Science 'Foundations Seismological Facilities
    ↪  for the Advancement of Geoscience (SAGE) Award under Cooperative Support
    ↪ Agreement OPP–1851037." )
Obligation( Include IRIS_PASSCAL_Polar, [ ], action = publish )

```
Attribute( IRIS_Trans, str "Data from the TA network were made freely available as part of the
    ↪ EarthScope USArray facility, operated by Incorporated Research Institutions for
    ↪ Seismology (IRIS) and supported by the National Science Foundation, under
    ↪ Cooperative Agreements EAR–1261681." )
Obligation( Acknowledge IRIS_Trans, [ ], action = publish )

Attribute( IRIS_PASSCAL_Mag, str "The magnetotelluric instruments were provided by the
    ↪ Incorporated Research Institutions for Seismology (IRIS) through the PASSCAL
    ↪ Instrument Center at New Mexico Tech. Data collected will be available through the
    ↪ IRIS Data Management Center. The facilities of the IRIS Consortium are supported by
    ↪ the National Science 'Foundations Seismological Facilities for the Advancement of
    ↪ Geoscience (SAGE) Award under Cooperative Support Agreement EAR–1851048." )
Obligation( Acknowledge IRIS_PASSCAL_Mag, [ ], action = publish )

Attribute( IRIS_Edu, str "Materials provided by the IRIS Education and Public Outreach Program
    ↪ have been used in this study. The facilities of the IRIS Consortium are supported by the
    ↪ National Science 'Foundations Seismological Facilities for the Advancement of
    ↪ Geoscience (SAGE) Award under Cooperative Support Agreement EAR–1851048." )
Obligation( Acknowledge IRIS_Edu, [ ], action = publish )

Attribute( IRIS_OBSIC, str "Data used in this research were provided by instruments from the
    ↪ Ocean Bottom Seismograph Instrument Center (obsic.who.edu) which is funded by the
    ↪ National Science Foundation. OBSIC data are archived at the IRIS Data Management
    ↪ Center ([url=http://www.iris.edu]http://www.iris.edu[/url]) which is funded by the
    ↪ National Science 'Foundations Seismological Facilities for the Advancement of
    ↪ Geoscience (SAGE) Award under Cooperative Support Agreement EAR–1851048." )
Obligation( Acknowledge IRIS_OBSIC, [ ], action = publish )
```

## F.11   OGL: Open Government Licence

This licence is stated on https://www.nationalarchives.gov.uk/doc/open-government-licence/version/3/. This is a general licence and each dataset may specify their own acknowledgment statement.

```
Attribute( OGL_ack, str "Contains public sector information licensed under the Open
    ↪ Government Licence v3.0." )
Obligation( Acknowledge OGL_ack, [ ], action = publish )
```

## F.12   World Bank

This policy is stated in https://www.worldbank.org/en/about/legal/terms-of-use-for-datasets. It contains nested policies, which refer to CC-BY and potential separate policies in its 3rd-party data. It explicitly re-specifies several aspects of CC-BY, so they are counted as a part of the policy.

Maybe because this policy is more close to the legal document, there is a large amount of disclaimer and contextual information. They constitute the general form of rules, but they are normally not actioning rules.

```
Obligation( Acknowledge WB, [ ], process = "publish" )
Attribute( WB, string "The World Bank: Dataset name: Data source (if known)" )
Obligation( Include CC_BY_SA_4 , [ ], process = "publish" )
```

Attribute( CC_BY_SA_4, url "https://creativecommons.org/licenses/by/4.0/" )

Obligation( Include WB_communicate, [ ], null)
Attribute( WB_communicate, str "If you have questions, seek to use Datasets on license terms
↪ other than the ones described above, or wish to make other comments, please contact
↪ us at +1 202 473 7824 or +1 800 590 1906, or by sending an email to data@worldbank.
↪ org". )